%% file: main.tex
\documentclass[fleqn,usenatbib]{mnras}

\usepackage{newtxtext,newtxmath}

\usepackage[T1]{fontenc}

\DeclareRobustCommand{\VAN}[3]{#2}
\let\VANthebibliography\thebibliography
\def\thebibliography{\DeclareRobustCommand{\VAN}[3]{##3}\VANthebibliography}

\usepackage{graphicx}	
\usepackage{amsmath}	

\newcommand{\oii}{[O~{\sc ii}]}
\newcommand{\oiii}{[O~{\sc iii}]}
\newcommand{\lya}{Ly$\alpha$}
\newcommand{\ha}{H$\alpha$}
\newcommand{\hb}{H$\beta$}

\newcommand{\fesc}{f$_{\rm{esc}}$}
\newcommand{\xion}{$\xi_{\rm{ion}}$}
\newcommand{\xionnofesc}{$\xi_{\rm{ion},0}$}
\newcommand{\ndot}{$\dot{n}_{\rm{ion}}$}
\newcommand{\Ndot}{$\dot{N}_{\rm{ion}}$}
\newcommand{\texp}{T$_{\mathrm{exp}}$}

\title[Ionising properties of galaxies in JADES]{Ionising properties of galaxies in JADES for a stellar mass complete sample: resolving the cosmic ionising photon budget crisis at the Epoch of Reionisation}

\author[C. Simmonds et al.]{
C. Simmonds,$^{1,2}$\thanks{E-mail: cs2210@cam.ac.uk}
S. Tacchella,$^{1,2}$
K. Hainline,$^{3}$
B.~D. Johnson,$^{4}$
D. Puskás,$^{1,2}$
B. Robertson,$^{5}$
\newauthor
W.~M. Baker,$^{1,2}$
R. Bhatawdekar,$^{6}$
K. Boyett,$^{7,8}$
A.~J. Bunker,$^{9}$
P.~A. Cargile,$^{4}$
S. Carniani,$^{10}$
\newauthor
J. Chevallard,$^{9}$
M. Curti,$^{11}$
E. Curtis-Lake,$^{12}$
Z. Ji,$^{3}$
G.~C. Jones,$^{9}$
N. Kumari,$^{13}$
I. Laseter,$^{14}$
\newauthor
R. Maiolino,$^{1,2,15}$
M.~V. Maseda,$^{14}$
P. Rinaldi,$^{3}$
A. Stoffers,$^{1,2}$
H. \"Ubler,$^{1,2}$
N.~C. Villanueva,$^{1,2}$
\newauthor
C.~C. Williams,$^{16}$
C. Willott,$^{17}$
J. Witstok,$^{1,2}$
and Y. Zhu$^{3}$ 
\\
$^{1}$The Kavli Institute for Cosmology (KICC), University of Cambridge, Madingley Road, Cambridge, CB3 0HA, UK\\
$^{2}$Cavendish Laboratory, University of Cambridge, 19 JJ Thomson Avenue, Cambridge, CB3 0HE, UK\\
$^{3}$Steward Observatory, University of Arizona, 933 N. Cherry Avenue, Tucson, AZ 85721, USA\\
$^{4}$Center for Astrophysics $|$ Harvard \& Smithsonian, 60 Garden St., Cambridge, MA 02138, USA\\
$^{5}$Department of Astronomy and Astrophysics University of California, Santa Cruz, 1156 High Street, Santa Cruz CA 96054, USA\\
$^{6}$European Space Agency (ESA), European Space Astronomy Centre (ESAC), Camino Bajo del Castillo s/n, 28692 Villanueva de la Cañada, Madrid, Spain\\
$^{7}$School of Physics, University of Melbourne, Parkville 3010, VIC, Australia\\
$^{8}$ARC Centre of Excellence for All Sky Astrophysics in 3 Dimensions (ASTRO 3D), Canberra 2611, Australia\\
$^{9}$Department of Physics, University of Oxford, Denys Wilkinson Building, Keble Road, Oxford OX1 3RH, UK\\
$^{10}$Scuola Normale Superiore, Piazza dei Cavalieri 7, I-56126 Pisa, Italy\\
$^{11}$European Southern Observatory, Karl-Schwarzschild-Strasse 2, 85748 Garching, Germany\\
$^{12}$Centre for Astrophysics Research, Department of Physics, Astronomy and Mathematics, University of Hertfordshire, Hatfield AL10 9AB, UK\\
$^{13}$AURA for European Space Agency, Space Telescope Science Institute, 3700 San Martin Drive. Baltimore, MD, 21210\\
$^{14}$Department of Astronomy, University of Wisconsin-Madison, 475 N. Charter St., Madison, WI 53706 USA\\
$^{15}$Department of Physics and Astronomy, University College London, Gower Street, London WC1E 6BT, UK\\
$^{16}$NSF’s National Optical-Infrared Astronomy Research Laboratory, 950 North Cherry Avenue, Tucson, AZ 85719, USA\\
$^{17}$NRC Herzberg, 5071 West Saanich Rd, Victoria, BC V9E 2E7, Canada\\
}

\date{Accepted XXX. Received YYY; in original form ZZZ}

\pubyear{2024}

\begin{document}
\label{firstpage}
\pagerange{\pageref{firstpage}--\pageref{lastpage}}
\maketitle

\begin{abstract}
We use NIRCam imaging from the JWST Advanced Deep Extragalactic Survey (JADES) to study the ionising properties of a sample of 14652 galaxies at $3 \leq z_{\rm{phot}} \leq 9$, 90\% complete in stellar mass down to log(M$_{\star}$/[M$_{\odot}$])$\approx 7.5$. Out of the full sample, 1620 of the galaxies have spectroscopic redshift measurements from the literature. 
We use the spectral energy distribution fitting code \texttt{Prospector} to fit all available photometry and infer galaxy properties. We find a significantly milder evolution of the ionising photon production efficiency (\xion\/) with redshift and UV magnitude than previously reported. Interestingly, we observe two distinct populations in \xion\/, distinguished by their burstiness (given by SFR$_{10}$/SFR$_{100}$). Both populations show the same evolution with $z$ and M$_{\rm{UV}}$, but have a different \xion\/ normalisation. We convolve the more representative $\log(\xi_{\rm{ion}} (z,\text{M}_{\rm{UV}}))$ relations (accounting for $\sim97$\% of the sample), 
 with luminosity functions from literature, to place constraints on the cosmic ionising photon budget. By combining our results, we find that one of our models can match the observational constraints from the \lya\/ forest at $z\lesssim6$. We conclude that galaxies with M$_{\rm{UV}}$ between $-16$ and $-20$, adopting a reasonable escape fraction, can produce enough ionising photons to ionise the Universe, without exceeding the required ionising photon budget.
\end{abstract}



\section{Introduction}
The Epoch of Reionisation (EoR) is one of the major phase transitions of the Universe, when it went from being dark and neutral to highly ionised, allowing Lyman Continuum (LyC; with energies above 13.6 eV) radiation to travel through the intergalactic medium (IGM). Observations place the end of this epoch at $z\sim6$ \citep[][]{Becker2001,Fan2006,Yang2020}, with some favouring a later reionisation at $z\sim5$ \citep{Keating2020,Bosman2022,Zhu2024}. There is a debate regarding which sources are the main responsible agents in ionising the Universe. The community widely agrees that young massive stars within galaxies are key, since they produce copious amounts of ionising photons, which might be able to escape the interstellar medium (ISM), and eventually ionise the IGM \citep{Hassan2018,Rosdahl2018,Trebitsch2020}. However, the nature of the galaxies that drive reionisation: bright and massive, faint and low-mass, or a combination of them, is still uncertain \citep{Finkelstein2019,Naidu2020,Robertson2022,Yeh2023}. Moreover, it is unclear how much active galactic nuclei (AGN) contribute to reionisation \citep{Dayal2020,Maiolino2023,Madau2024}. 

The stellar mass of galaxies has been seen to correlate with how efficiently ionising photons are produced \citep{Simmonds2023_JADES}. Simulations indicate that it also relates how these ionising photons escape \citep{Paardekooper2015}. The latter is measured through their Lyman Continuum escape fractions (\fesc\/(LyC)), defined as the ratio between the H-ionising radiation that is emitted intrinsically, and that which reaches the IGM. In order for galaxies to account for the reionisation of the Universe, either a significant average \fesc\/ value is required \citep[\fesc\/=10-20\%;][]{Ouchi2009,Robertson2013,Robertson2015,Finkelstein2019,Naidu2020}, or a high ionising photon production efficiency. These ranges of \fesc\/ have been observed for individual star-forming galaxies at $z \lesssim 4$ \citep[e.g. ][]{Borthakur2014,Bian2017,Vanzella2018,Izotov2021}, but not usually in large numbers \citep[][]{Leitet2013,Leitherer2016,Steidel2018,Flury2022a}. An alternative to high escape fractions is a high ionising photon production efficiency (\xion\/), given by the ratio between the rate of ionising photons being emitted (\ndot\/), and the monochromatic non-ionising ultra-violet (UV) luminosity density. Indeed, observational studies up to $z\sim 9$ have found \xion\/ to increase as a function of redshift \citep[e.g. ][]{Endsley2021,Stefanon2022,Tang2023,Atek2023,Simmonds2023_JEMS,Simmonds2023_JADES,Harshan2024,Pahl2024,Saxena2024}.

The behaviour of \xion\/ as a function of redshift has important consequences on the cosmic budget of reionisation \cite[e.g. ][]{Munnoz2024}, defined as the number of ionising photons produced per comoving volume unit of the Universe (\Ndot\/). Three ingredients must be provided in order to study \Ndot\/: (1) a prescription for \fesc\/, (2) a UV luminosity density function, $\rho_{\rm{UV}}$, describing the number of galaxies per unit volume that have a given UV luminosity, as a function of redshift \citep[e.g. ][Whitler et al. in prep.]{Bouwens2021,Adams2024, Donnan2024, Robertson2024}, and (3) constraints on \xion\/. In addition, the IGM clumping factor of the Universe has to be considered \citep[e.g. ][]{Madau1999,Kaurov2014,So2014}. This factor is a measure of the uniformity of the matter distribution in the Universe, and has crucial implications on reionisation since it relates to the amount of atomic recombinations taking place in the IGM. Briefly, a higher clumping factor implies that more ionising photons need to be emitted per unit volume at a given redshift in order to sustain hydrogen ionisation. 

Before the launch of the James Webb Space Telescope \citep[JWST; ][]{Gardner2023}, it was common practice to set \fesc\/ and \xion\/ as constants. Fortunately, the JWST has given us an unprecedented view of the early Universe at restframe optical wavelengths, which has allowed us to place better constraints on \xion\/. In particular, \cite{Simmonds2023_JADES} studied a sample of emission-line galaxies (ELGs) at $z\sim 4-9$ using photometry obtained with the Near Infrared Camera \citep[NIRCam; ][]{Rieke2023}, on board the JWST. The mean redshift of reionisation is at $z = 7.68\pm 0.79$ \citep{Planck2020}, meaning that this study (and the many others enabled by JWST) probe deep into the EoR. Through \ha\/ and \oiii\/ emission line fluxes, \cite{Simmonds2023_JADES} estimated \xion\/ for a sample of 677 galaxies. In parallel, they inferred the same quantity by using the spectral energy distribution (SED) fitting code \texttt{Prospector} \citep{Johnson2019,Johnson2021}. They find that the \xion\/ measurements estimated by emission line fluxes agree with the values obtained by \texttt{Prospector}. Additionally, they conclude that  \xion\/ increases with redshift, and that this increase is due to low-mass faint galaxies having more bursty star formation histories (SFHs). Here burstiness is quantified by the ratio between recent (averaged over 10 Myr) and past (averaged over 100 Myr) star formation rates (SFRs), which is associated with low stellar masses \citep{Weisz2012,Guo2016,Looser2023a}, mainly due to the increased importance of stellar feedback. At high redshifts, however, burstiness can also be explained by the imbalance between gas accretion and supernovae (SNe) feedback time scales, which prevent star formation equilibrium in the ISM \citep{Faucher-Giguere2018,Tacchella2020}. We note that the SFR$_{10}$/SFR$_{100}$ ratio is a direct measure of the recent SFH, and its variance for an ensemble of galaxies measures short-term star formation variability \citep[``burstiness'';][]{Caplar2019}.

The sample constructed in \cite{Simmonds2023_JADES} suffered from one main limitation: since emission lines were required, and with a sufficient strength so that they were measurable from photometry, this sample was biased towards star-forming galaxies with significant \ha\/ and/or \oiii\/ emission. In fact, Laseter et al. (in prep.) demonstrate \xion\ is consistently high (log(\xion\//[Hz erg$^{-1}$]) $\approx 25.5$) down to \oiii\/ equivalent widths of 200\AA\/ due to low metallicities ($Z \lesssim 1/10 Z_{\odot}$), further demonstrating this bias from the past results. Given the agreement found between \texttt{Prospector} and the emission line measurements of \xion\/, in this work, we use \texttt{Prospector} to fit the full JWST Advanced Deep Extragalactic Survey \citep[JADES; ][]{Eisenstein2023, Bunker2023} photometry set for a sample of JADES galaxies in the Great Observatories Origins Deeps Survey South \citep[GOODS-S; ][]{Giavalisco2004}. Our sample is 90\% complete in stellar mass down to masses of log(M$_{\star}$/[M$_{\odot}$])$\sim 7.5$, providing us a deep statistical view of the ionising properties of galaxies.

The structure of this paper is the following. In $\S$~\ref{sec:data} we present the data used in this work, along with the sample selection criteria. In $\S$~\ref{sec:Prospector}  we present our \texttt{Prospector} fitting method. Some general properties of the sample are given in $\S$~\ref{sec:general_properties}, followed by our constraints on the ionising properties of galaxies in  $\S$~\ref{sec:ionising_properties}. Implications for reionisation are discussed in $\S$~\ref{sec:implications_reionisation}, while the caveats and limitations of our methods are discussed in $\S$~\ref{sec:caveats}. Finally, brief conclusions are presented in $\S$~\ref{sec:conclusions}. Throughout this work we assume $\Omega_0 = 0.315$ and $H_0 = 67.4$ km s$^{-1}$ Mpc$^{-1}$, following \cite{Planck2020}. 

\begin{figure*}
	\includegraphics[width=2\columnwidth]{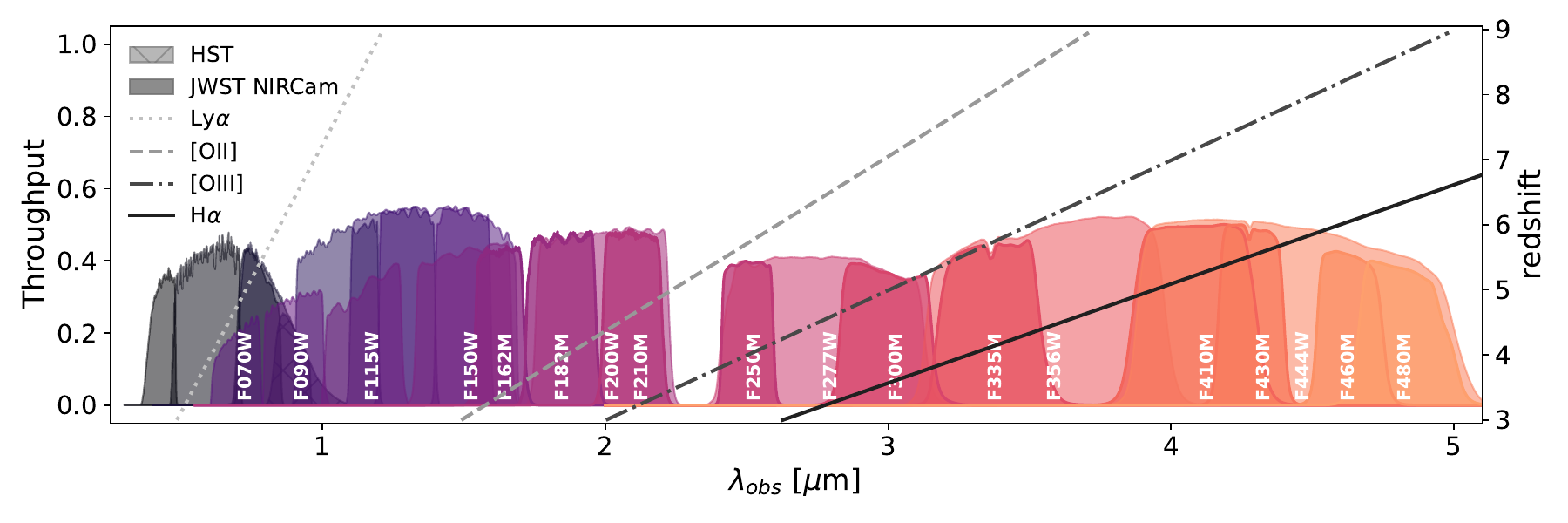}
 \includegraphics[width=2\columnwidth]{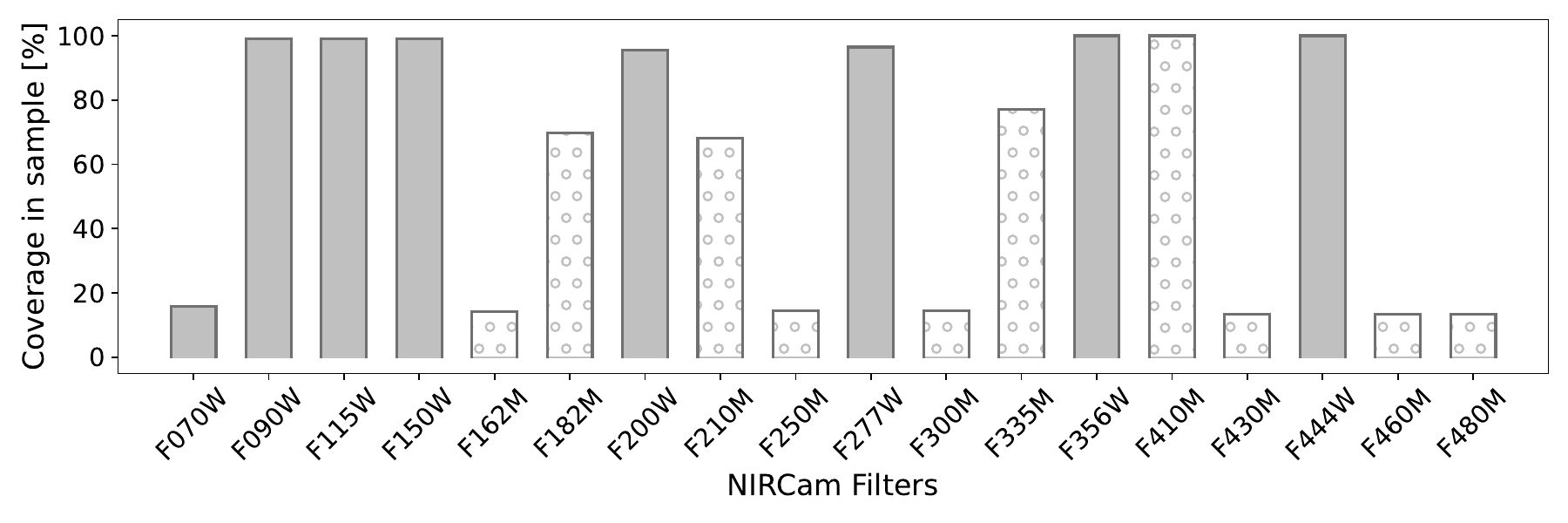}
    \caption{\textsl{Top panel:} Throughputs of the HST and NIRCam filters used in the SED fitting in this work for galaxies at $3\leq z_{\rm{phot}} \leq 9$. It is important to note that not all of these filters are available for every galaxy studied in this work. The hatched areas show the HST ACS bands: F435W, F606W, F775W, F814W, F850LP, and the HST IR bands: F105W, F125W, F140W, and F160W. While the filled regions are labelled and show the JWST NIRCam bands, from left to right: F070W, F090W, F115W, F150W, F162M, F182M, F200W, F210M, F250M, F277W, F300M, F335M, F356W, F410M, F430M, F444W, F460M, and F480M. The lines show the observed wavelengths of selected emission lines (\lya\/, \oii\/, \oiii\/, and \ha\/) with redshift. \textsl{Bottom panel:} percentage of sources in our final, stellar mass-complete sample that are covered by each JWST NIRCam filter. The medium bands are shown as dotted areas.}
    \label{fig:filter_curves}
\end{figure*}

\begin{figure*}
    \includegraphics[width=2\columnwidth]{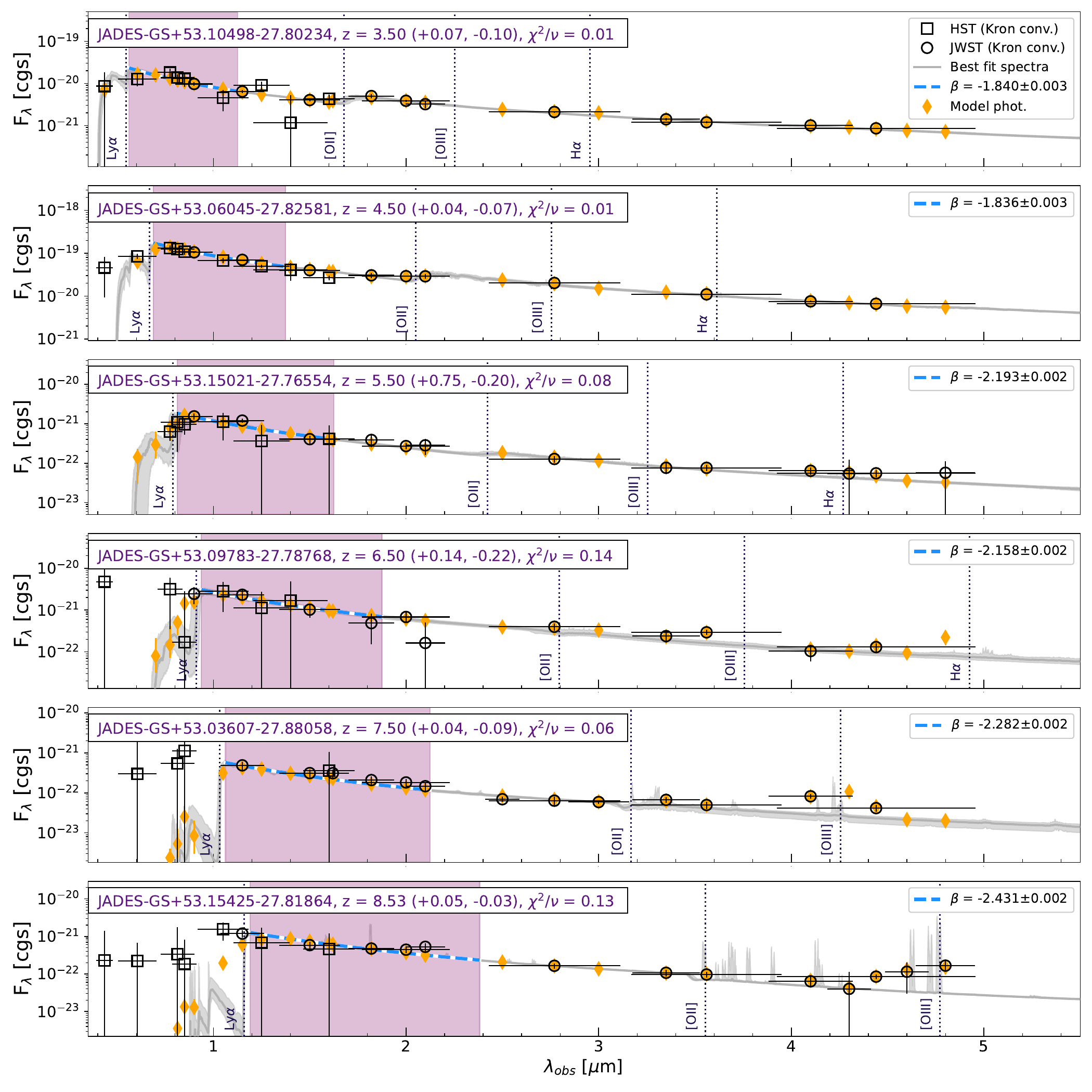}
    \caption{Representative example SEDs and best-fit spectra for galaxies in our sample, assuming a continuity (i.e. non-parametric) SFH. From top to bottom the redshift increases from $z = 3.5$ to 8.5, in steps of 1, the galaxy identifiers, along with their $z_{\rm{phot}}$ and reduced $\chi^2$ (for the JWST NIRCam photometry), are shown in the top left corner of each panel. The symbols show the photometric points for HST (open squares), JWST NIRCam (open circles), and model photometry (orange diamonds), respectively. The grey curves show the best-fit spectra obtained by \texttt{Prospector}, with the spectral region used to estimate UV continuum slope ($\beta$; $\lambda_{\rm{rest-frame}} = 1250 - 2500$ \AA\/) shaded in purple. The observed wavelengths of \lya\/, \oii\/, \oiii\/, and \ha\/ are shown as vertical dotted lines.}
    \label{fig:SED_examples}
\end{figure*}

\begin{figure}
	\includegraphics[width=\columnwidth]{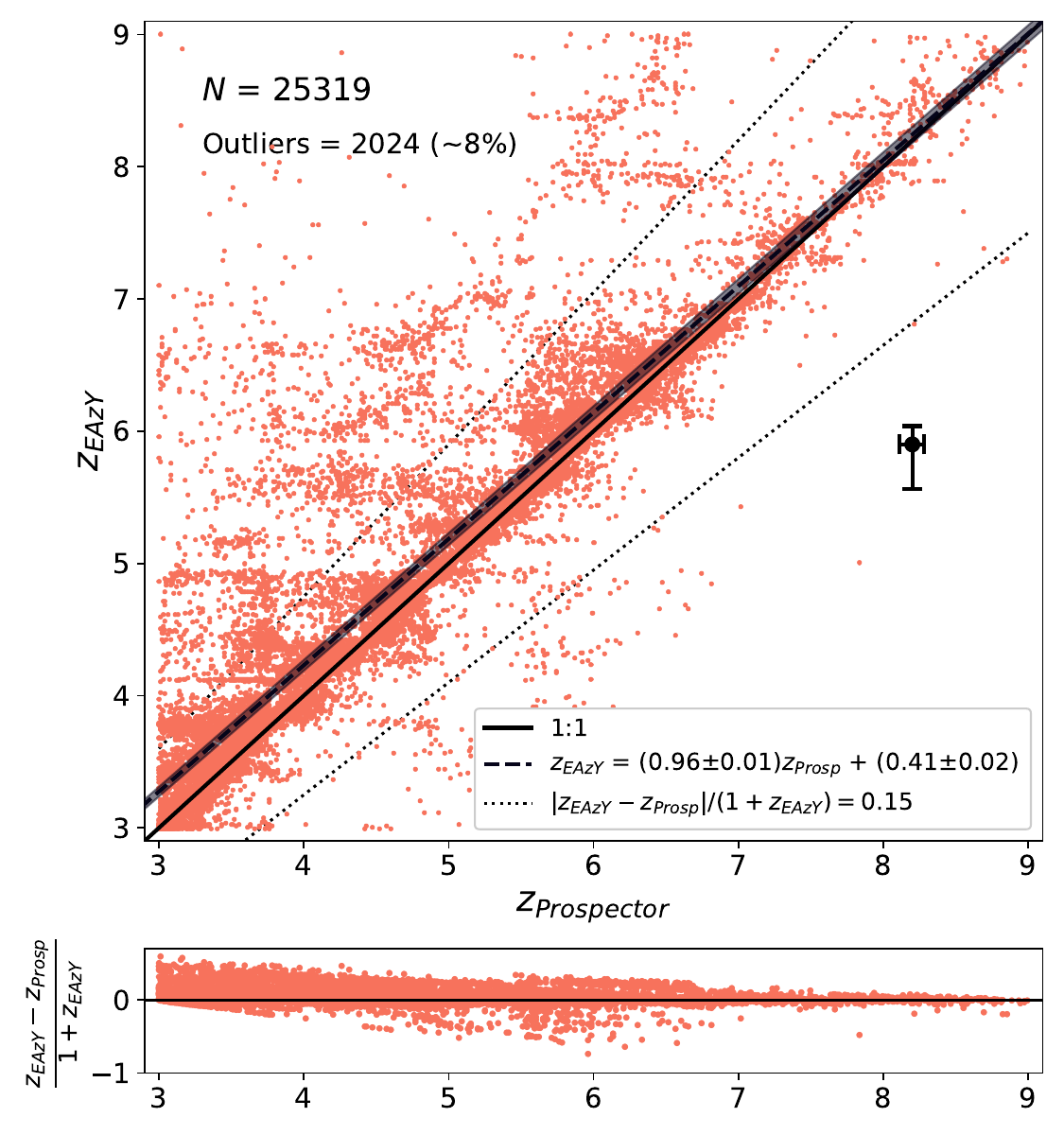}
    \caption{Comparison of input and output photometric redshifts. The vertical axis shows the photometric redshifts inferred using \texttt{EAzY}. These redshifts were used as priors in the SED fitting with \texttt{Prospector}, the medians of these posteriors are shown in the horizontal axis. The black point shows the redshift median errors, we note that the \texttt{Prospector}-inferred values tend to be better constrained, but that they overall follow a one-to-one trend with those inferred using \texttt{EAzY}, as can be seen by the best-fit line (dashed black line). The points that fall outside of the dotted black lines are considered catastrophic outliers, there are 2036 galaxies that fall into this category, corresponding to $\sim 8\%$. We note that the outlier fraction is estimated by comparing the median values inferred by \texttt{EAzY} and \texttt{Prospector}. The former has considerably larger error bars in general.}
    \label{fig:photoz_comparisons}
\end{figure}

\begin{figure}
    \includegraphics[width=\columnwidth]{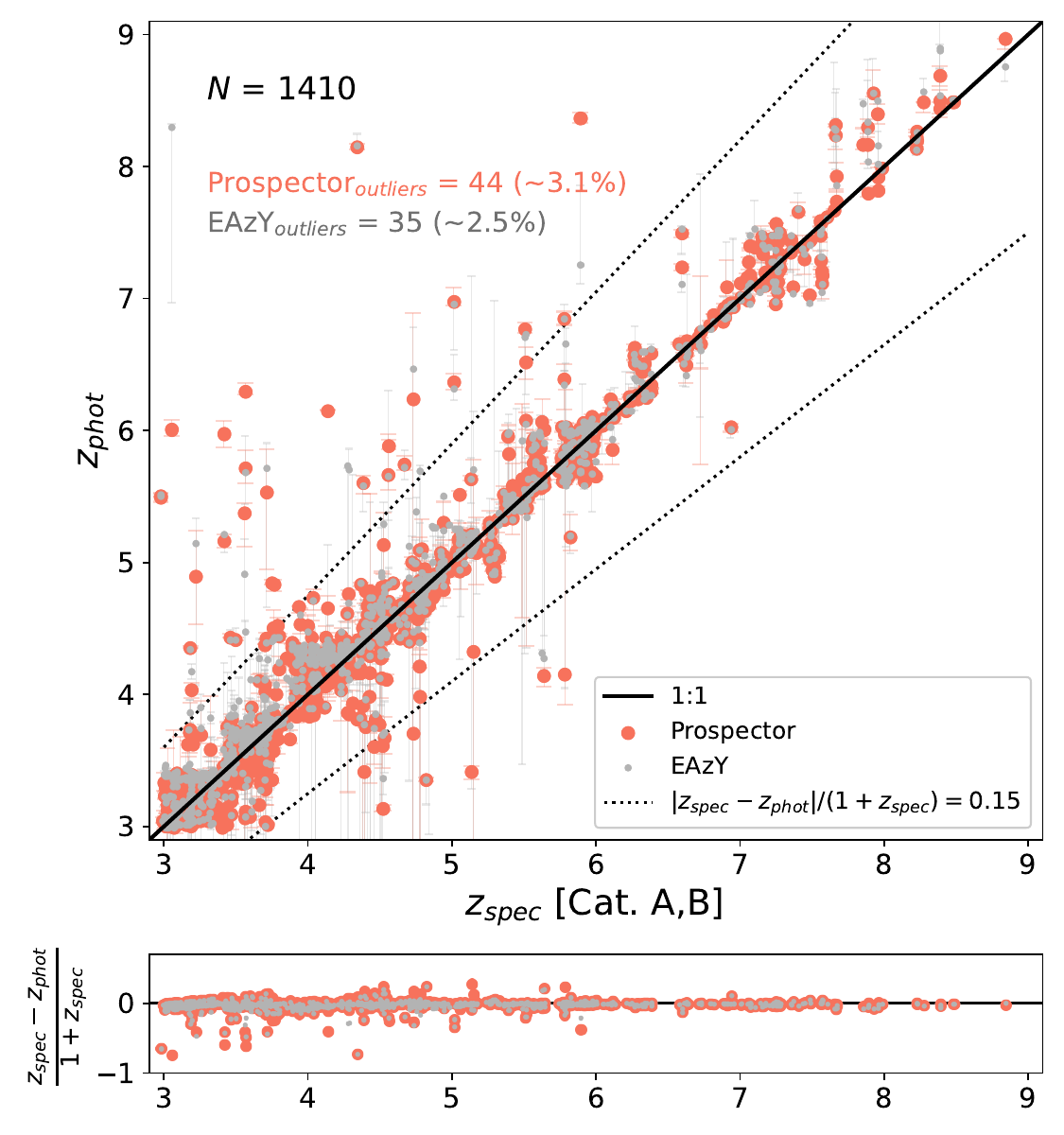}
    \caption{Comparison of photometric and spectroscopic redshifts, when spectroscopic redshifts in category A and/or B are available. Both \texttt{Prospector} (larger orange circles) and \texttt{EAzY} (smaller grey circles) inferred values generally follow a one-to-one relation with the spectroscopic redshifts, with some exceptions. We conclude that both codes can retrieve $z_{\rm{spec}}$ successfully in the majority of cases. The points that fall outside of the area delimited by the dotted lines are considered catastrophic outliers. There are 44 (35) of such objects inferred by \texttt{Prospector} (\texttt{EAzY}), corresponding to $\sim 3.1\%$ ($\sim$ 2.5\%) of the subsample with spectroscopic redshifts in categories A and B.}
    \label{fig:photoz_specz_comparisons}
\end{figure}

\begin{figure}
	\includegraphics[width=\columnwidth]{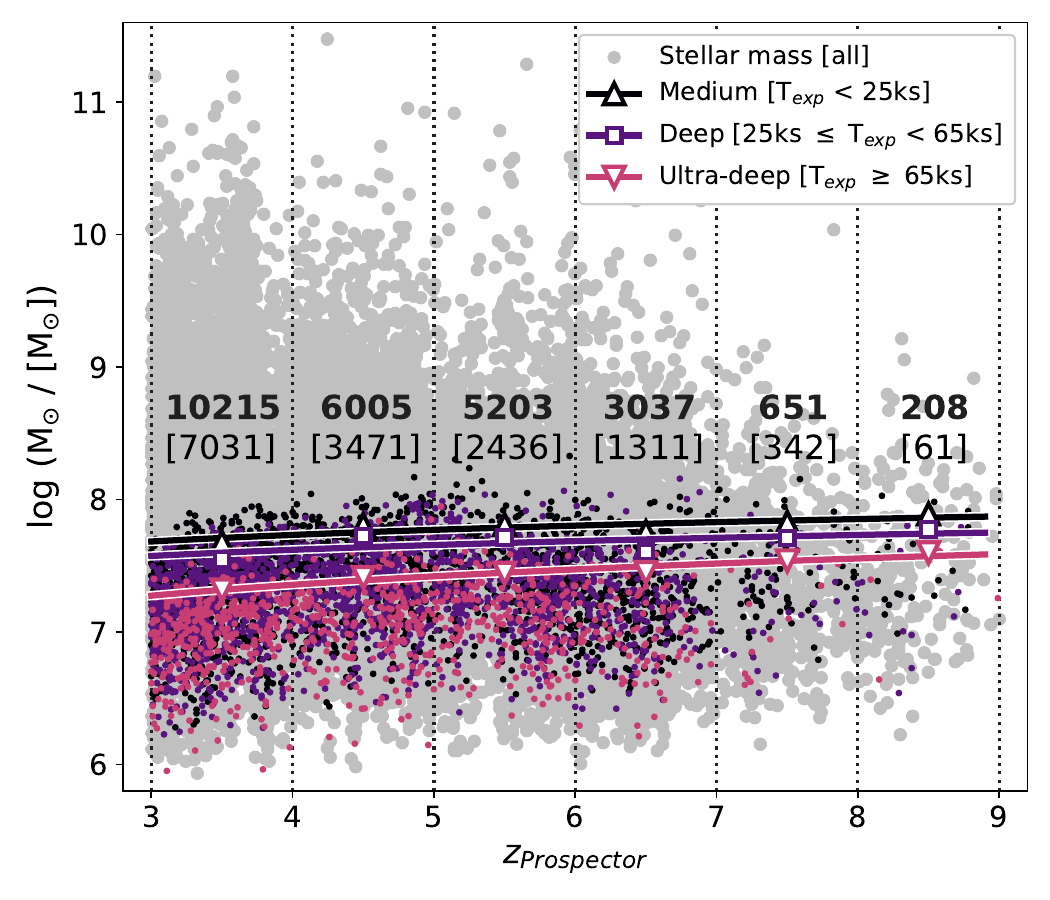}
    \caption{Stellar mass completeness of the sample, divided into three fields depending on exposure time (\texp\/). The larger grey circles show the stellar mass of each galaxy as a function of redshift, while the smaller coloured circles show the limiting mass for the faintest 20\%. The solid curves denote the 90\% completeness at each redshift and for each depth, as indicated in the legend, following the prescription of \citet{Pozzetti2010}. We find our sample is stellar mass complete down to log(M$_{\star}$/[M$_{\odot}$]) $\approx 7.5$, with small variations depending on the depth of the field. The dotted vertical lines mark the redshift bins used in the mass completeness estimation, while the numbers provided in each bin indicate the total number of galaxies in the bin (top row, bold faced), and the galaxies above the limiting mass (bottom row, in brackets).}
    \label{fig:completeness}
\end{figure}


\begin{figure*}
\includegraphics[width=2\columnwidth,trim={7cm 2cm 7cm 0}]{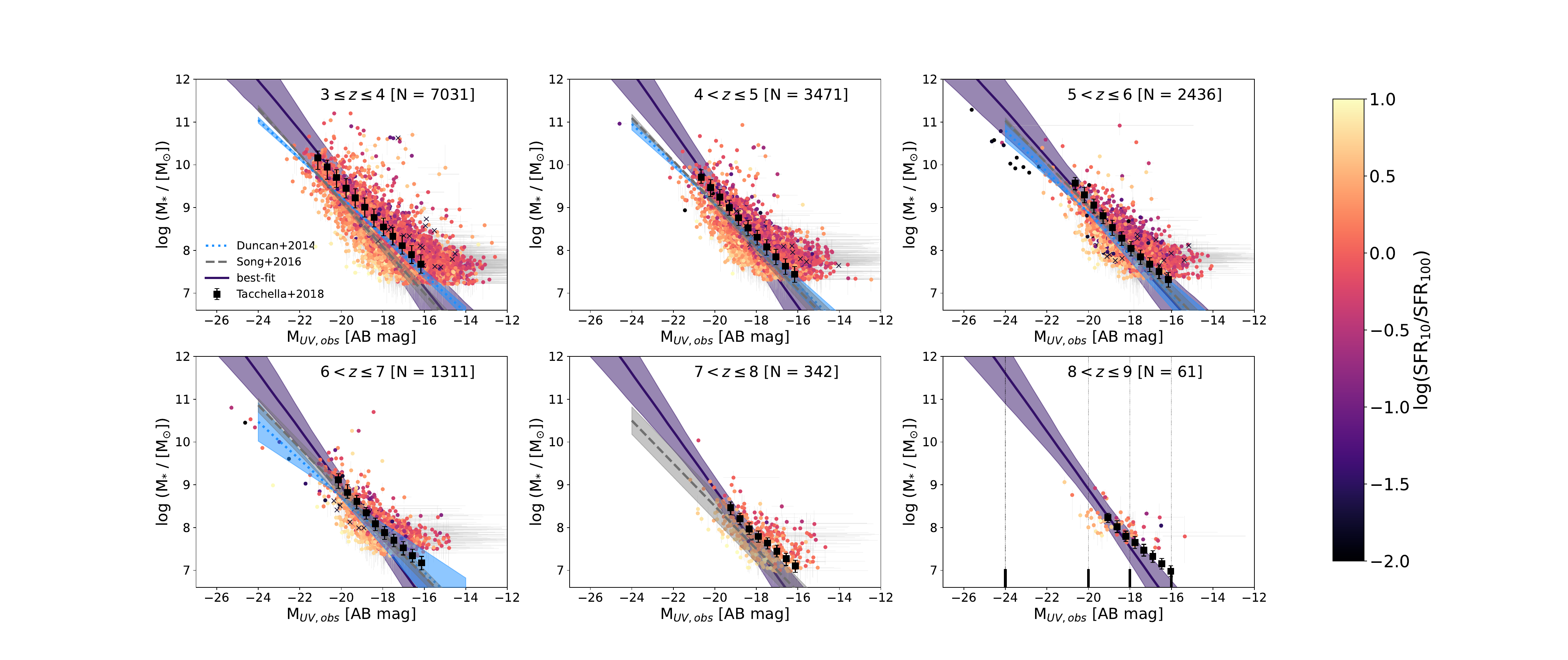}
    \caption{Stellar mass as a function of (observed) M$_{\rm{UV}}$ for our stellar mass complete sample, divided into redshift bins, and colour-coded by the burstiness of their SFHs (i.e. ratio between star formation in the past 10 Myr and the one averaged over the last 100 Myr). For comparison, we overlay the log(M$_{\star}$)-M$_{\rm{UV}}$ mass-to-light relations from \citet{Duncan2014}, \citet{Song2016}, and \citet{Tacchella2018}. The best-fit relation, using forward modelling to take account for the completeness of the sample, is shown as a purple solid curve and shaded area (see Table~\ref{tab:bestfit_M_MUV}). The black vertical lines in the bottom right panel show the limits of three M$_{\rm{UV}}$ bins delimited by M$_{\rm{UV}} = -24, -20, -18$, and $-16$, which are used later in Figure~\ref{fig:cosmic_Ndot_percentages}. As expected, brighter galaxies tend to have higher stellar masses, however, there is a considerable scatter in M$_{\rm{UV}}$ for a fixed stellar mass.}
    \label{fig:MUV_mass_zbins}
\end{figure*}

\begin{figure*}
    \includegraphics[width=2\columnwidth]{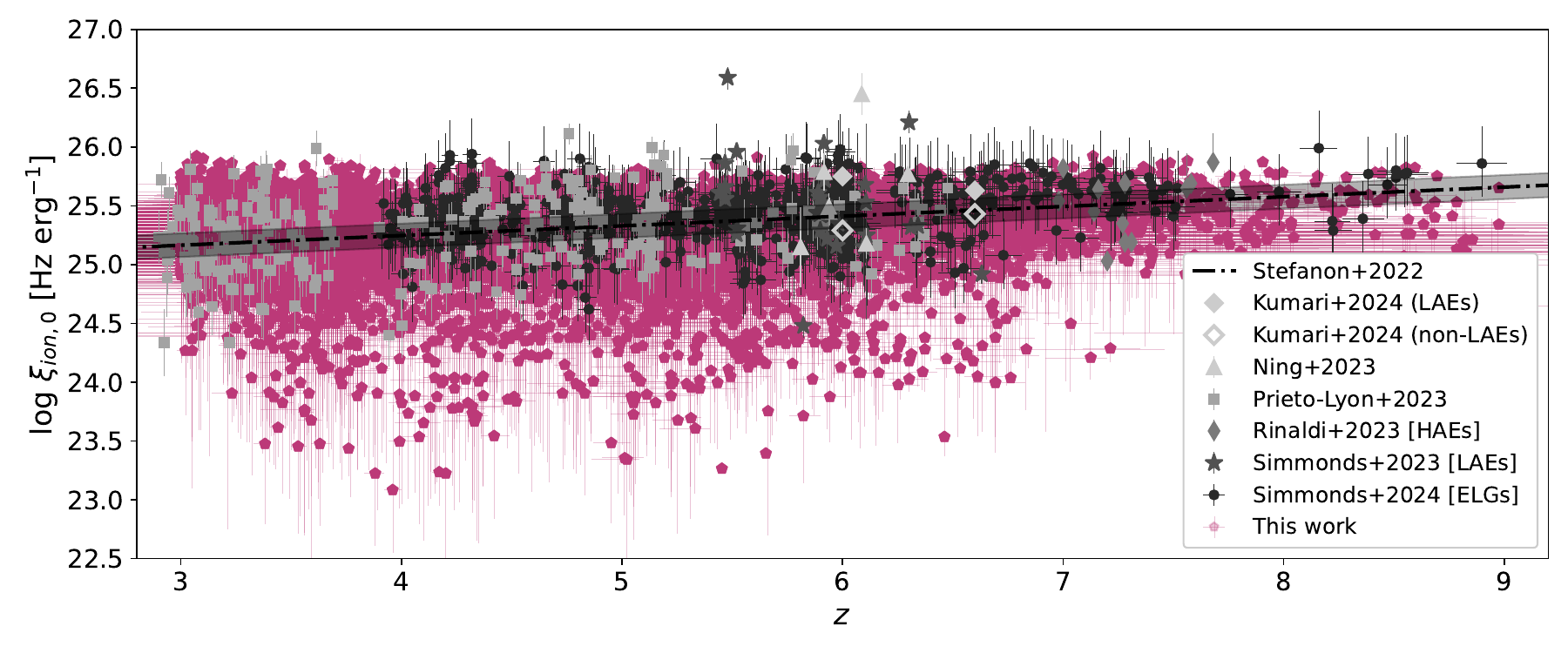}
    \caption{Compilation of \xionnofesc\/ from literature, in comparison to our stellar mass complete sample (this work, pentagons). The best-fit line from \citet{Stefanon2022}, with a slope of dlog(\xionnofesc\/)/dz $= 0.09 \pm 0.01$, is shown as a black dashed line and shaded area. In order of increasing redshift, the markers in grey scale represent: UV-faint galaxies at $z \sim 3-7$ from \citet{Prieto-Lyon2023} (squares), \lya\ emitters from \citet{Ning2023} (triangles) and \citet{Simmonds2023_JEMS} (stars),  the \ha\ emitters from \citet{Rinaldi2024HAEs} (thin diamonds), and finally, the ELGs from \citet{Simmonds2023_JADES}, which have a slope of dlog(\xionnofesc\/)/dz $= 0.07 \pm 0.02$. The thick diamonds show the stacked results from \citet{Kumari2024}, for galaxies above and below $z = 6.3$. Due to observational limitations and the emission line methods used to estimate \xionnofesc\/ (i.e.  through \ha\/ and/or \oiii\/ emission) the previous samples were biased towards star-forming galaxies. Our stellar mass complete sample reveals a population of galaxies with low ionising photon production efficiency (log(\xionnofesc\//[Hz erg$^{-1}$])$\lesssim$ 24.5), and a considerably less significant evolution with redshift (dlog(\xionnofesc\/)/dz $\sim 0.01$, see Figure~\ref{fig:xion_nion_redshift}).}
    \label{fig:xion_literature}
\end{figure*}

\begin{figure*}
    \includegraphics[width=2\columnwidth,trim={3cm 0 3cm 0}]{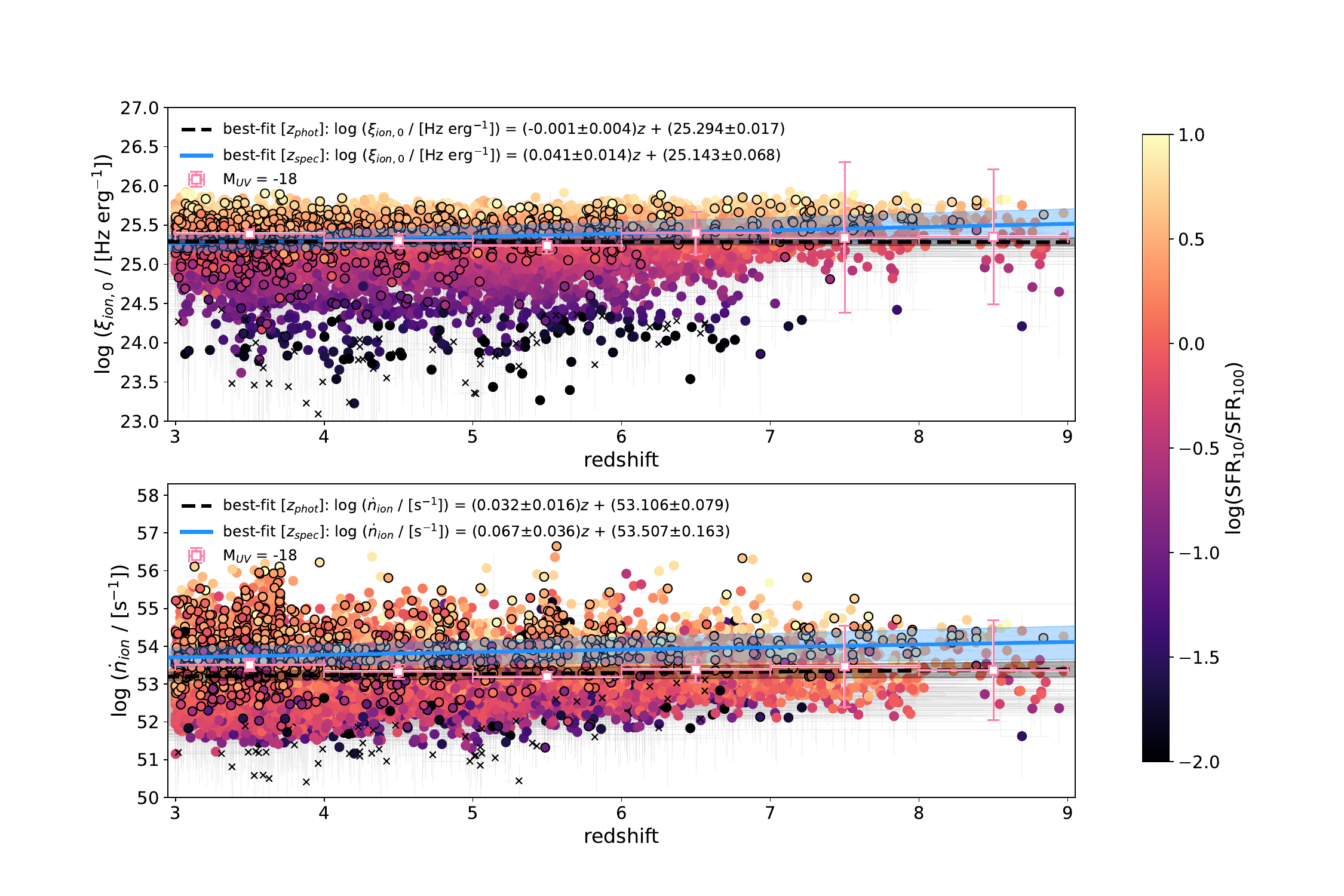}
    \caption{\xionnofesc\/ (top) and \ndot\/ (bottom) inferred through SED fitting with \texttt{Prospector}, as a function of redshift, colour-coded by a measure of burstiness (defined here as the ratio between the star formation sustained in the past 10 Myr and the one averaged over the past 100 Myr), for the stellar mass complete sample (log(M$_{\star}$/[M$_{\odot}$])$> 7.5$). The circles with black edges show the spectroscopic sample, while the circles without edges show the photometric sample. Within the latter, the galaxies with zero recent star formation (SFR$_{10}$ = 0 M$_{\odot}$ yr$^{-1}$) are shown as crosses. The best-fit relations to both samples are shown in the top left corner of each panel. The best fit to the spectroscopic data (blue-filled line) yields a more positive slope than the one derived for the entire photometric sample (black dashed line). Particularly, the fit to \xionnofesc\/ in the spectroscopic sample (dlog(\xionnofesc\/)/dz $\sim 0.05\pm 0.02$) is closer to the \xionnofesc\/ versus redshift relations from the literature \citep[within errors of the findings of ][]{Simmonds2023_JADES}. This is because the spectroscopic sample likely suffers from the same biases as previous studies (i.e.  biased towards star-forming galaxies with detectable emission lines). Finally, the pink squares with error bars show the best-fit relations per redshift bin shown in Figures~\ref{fig:xion_MUV_zbins} and~\ref{fig:nion_MUV_zbins}, for a fixed M$_{\rm{UV}}$ of -18. When the full stellar mass complete sample is considered, both \xionnofesc\/ and \ndot\/ show only a slight evolution with redshift, \xionnofesc\/ shows a strong correlation with burstiness, while \ndot\/ does not.}
    \label{fig:xion_nion_redshift}
\end{figure*}

\begin{figure*}    \includegraphics[width=2.1\columnwidth,trim={7cm 2cm 7cm 0}]{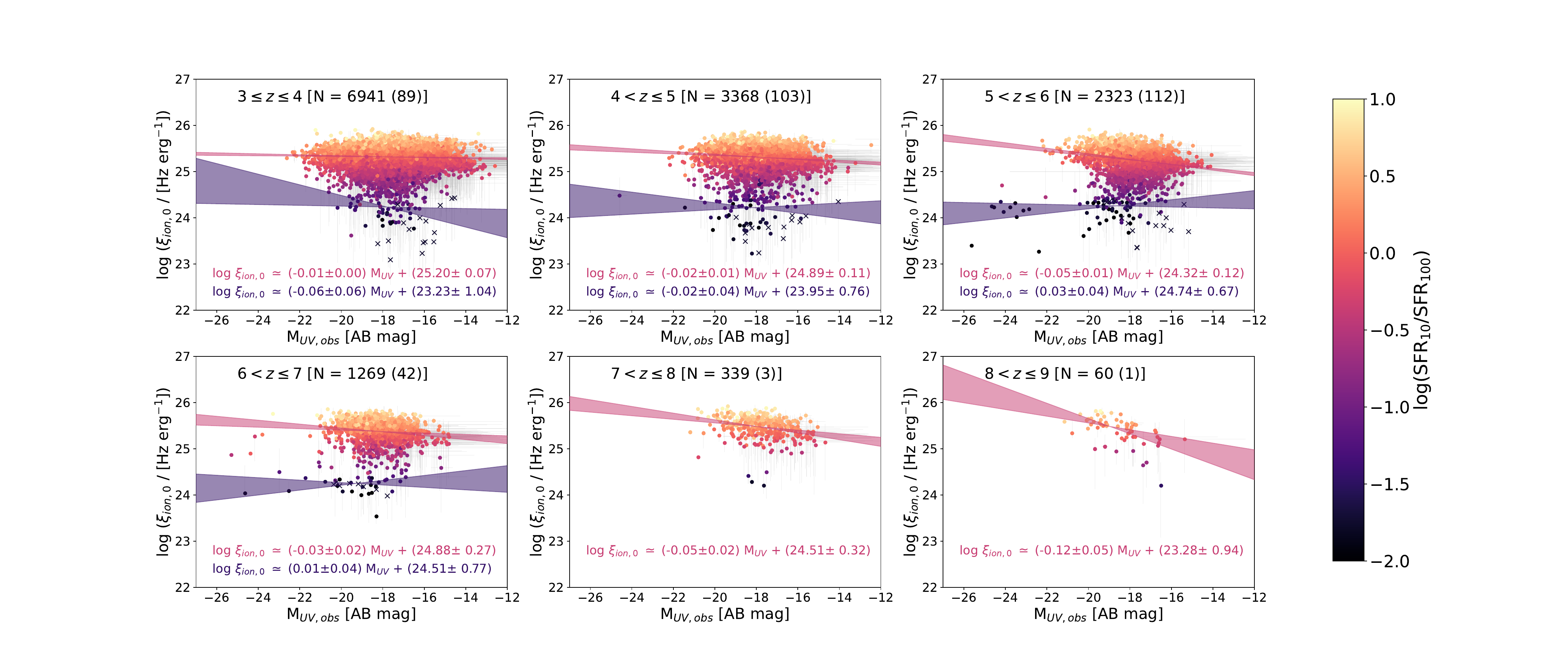}
    \caption{Dependence of \xionnofesc\/ on UV magnitude, separated in redshift bins, and colour-coded by the burstiness of their SFHs. The individual error bars are shown in grey. The top left of each panel shows the redshift range and the amount of galaxies it contains. The number in parenthesis corresponds to a small secondary population of galaxies with log(SFR$_{10}$/SFR$_{100}$) $< -1$, that lie systematically below the general \xionnofesc\ trends with M$_{\rm{UV}}$ and redshift. If these two populations of galaxies are fit separately (when possible), the slope of \xionnofesc\/ with M$_{\rm{UV}}$ is consistent between them, however, their \xionnofesc\/ intercept is different. The values for both fits are shown at the bottom of each panel, the darker coloured shaded area and text correspond to the population with no (or very little) recent star formation. We note that this population accounts for only $< 3$\% of the total sample.}
    \label{fig:xion_MUV_zbins}
\end{figure*}

\begin{figure*}
    \includegraphics[width=2.1\columnwidth,trim={7cm 2cm 7cm 0}]{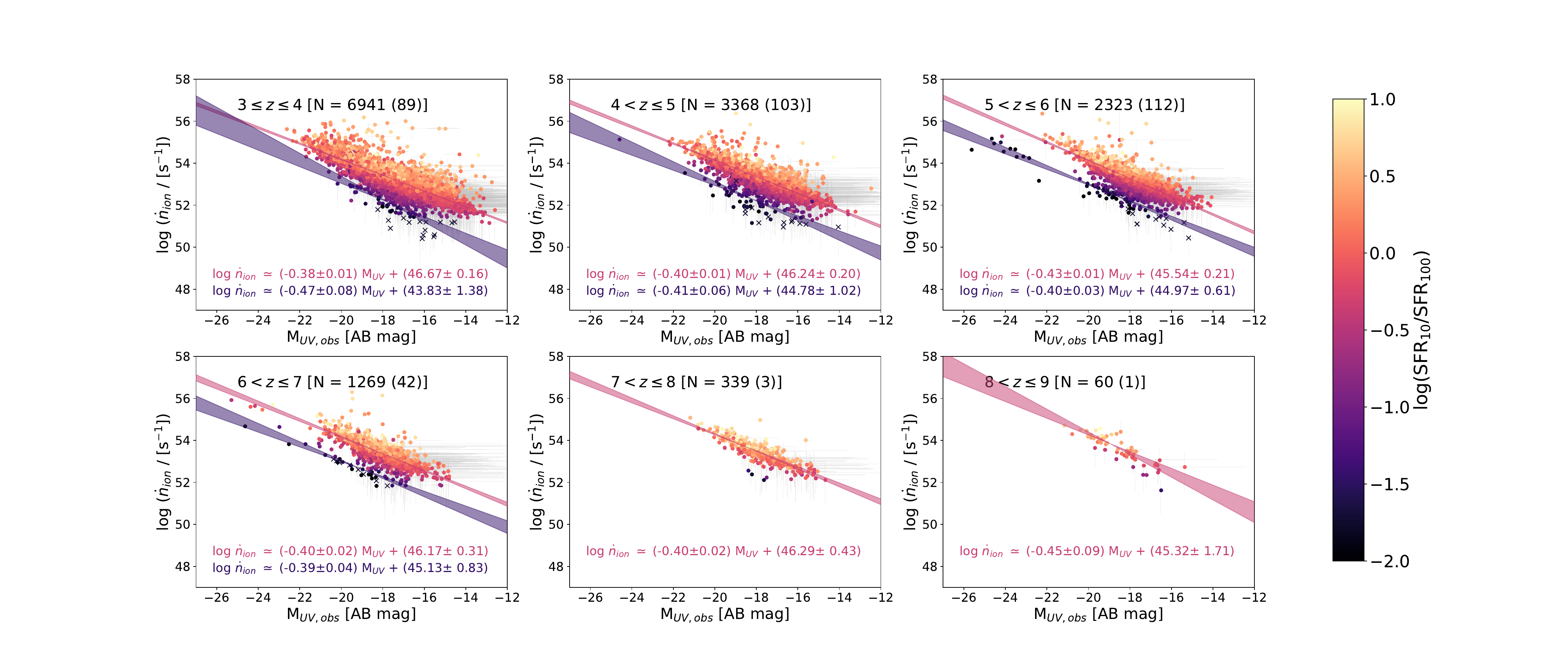}
    \caption{Same as Figure~\ref{fig:xion_MUV_zbins} but for \ndot\/ instead. Contrary to the case of \xionnofesc\/, \ndot\/ has a stronger dependence on M$_{\rm{UV}}$, where the faintest (brightest) galaxies produce less (more) ionising photons.}    
    \label{fig:nion_MUV_zbins}
\end{figure*}

\section{Data and selection criteria}
\label{sec:data}
In this section, we describe the data and selection criteria applied to build our sample of galaxies, with the goal to construct a stellar mass complete sample between redshifts 3 and 9. 

\subsection{Data}
We use the full JADES \citep[][]{Eisenstein2023, Bunker2023} photometry set in the GOODS-S region, including the publicly available NIRCam Deep imaging \citep{Rieke2023}, and the JADES Origins Field \citep[JOF; ][]{Eisenstein2023JOF}, covering an area of $\sim 45$ arcmin$^2$ with an average exposure time of 130 hours. When available, we also use photometry from the JWST Extragalactic Medium-band Survey \cite[JEMS; ][]{Williams2023}, and from the First Reionisation Epoch Spectroscopic Complete Survey \citep[FRESCO; ][]{Oesch2023}. 

\subsubsection{Photometry}
The photometric catalogue used in this work has been produced in the same way as the one used in \cite{Simmonds2023_JADES}. In brief, the source detection and photometry leverage both the JEMS NIRCam medium band and JADES NIRCam broad and medium band imaging. Detection was performed using the \texttt{photutils} \citep{bradley2022a} software package, identifying sources with contiguous regions of the SNR mosaic with signal $>3\sigma$ and five or more contiguous pixels. We also use {\tt photutils} to perform circular aperture photometry with filter-dependent aperture corrections based on model point-spread-functions following the method of \cite{Ji2023}, as described in \cite{Robertson2024}. In addition to the NIRCam observations, HST images from the Hubble Legacy Field programme \citep{Illingworth2016, Whitaker2019}, and the Cosmic Assembly Near-Infrared Deep Extragalactic Legacy Survey \citep[CANDELS; ][]{Grogin2011, Koekemoer2011}, as well as reductions of GOODS-S from \cite{Giavalisco2004}, are used. The details of the catalogue generation and photometry will be presented in Robertson et al., (in prep). In this work, we use a Kron aperture placed on images that have been convolved to a common resolution, and impose an error floor of 5\% in each band. Our photometry does not take include the \texttt{EAzY} derived photometric offsets, however, we find that the NIRCam photometric offsets are small given the uncertainty of the photometry (see Appendix~\ref{app:chi}). The throughputs of the filters used in this work are shown in Figure~\ref{fig:filter_curves}, as well as emission lines of interest and how their observed wavelength evolves with redshift.

\subsubsection{Redshifts}
Due to the richness of the photometry set, photometric redshifts ($z_{\rm{phot}}$) can be obtained with great accuracy \citep{Hainline2023}. When fitting SEDs in this work, we use the $z_{\rm{phot}}$ inferred by the template-fitting code \texttt{EAzY} \citep{Brammer2008}, as described in \cite{Hainline2023} and \cite{Rieke2023}.  Additionally, when available, we make use of spectroscopic redshifts ($z_{\rm{spec}}$) from the Near-Infrared Spectrograph \cite[NIRSpec; ][]{Jakobsen2022}, as well as those reported in literature (Pusk\'as et al. in prep), including the JADES NIRSpec redshifts from \cite{Bunker2023} and \cite{D'Eugenio2024}. In particular, they have been compiled from the Atacama Large Millimeter/submillimeter Array (ALMA) Spectroscopic Survey in the Hubble Ultra-Deep Field \citep[ASPECS; ][]{Walter2016}, CANDELS \citep{Grogin2011}, the 3D-Hubble Space Telescope (HST) Survey \citep{Momcheva2016}, the Multi Unit Spectroscopic Explorer \citep[MUSE; ][]{Bacon2010} Ultra-Deep Field DR2 \citep{Bacon2022}, and a redshift catalogue (F. Sun private communication) produced from grism data of the First Reionisation Epoch Spectroscopic Complete Survey \citep[FRESCO; ][]{Oesch2023}.  
 
 We selected galaxies that were matched to JADES NIRCam positions within 0.25 arcsec, and that were flagged as reliable by each team. Since galaxies often have multiple $z_{\rm{spec}}$ measurements, we defined four categories to collate the several redshift measurements into one, which we call $z_{\rm{best}}$:  
\begin{itemize}
    \item \textbf{Category A:} Only one redshift labelled as having the highest quality. We use this redshift as $z_{\rm{best}}$.
    \item \textbf{Category B:} Multiple redshifts with the same (highest) quality that agree when rounded to the second decimal. We use this rounded value as $z_{\rm{best}}$.
    \item \textbf{Category C:} Multiple solutions with the same (highest) quality that have a non-dramatic disagreement (with a difference smaller than 1). We define $z_{\rm{best}}$ as the mean between the solutions and add errors reflecting the difference.
    \item \textbf{Category D:} Multiple solutions with the same (highest) quality that have a dramatic disagreement ($\Delta > 1$). In these cases, we keep all the highest quality redshifts and follow up with a visual inspection.
\end{itemize}

\subsection{Sample selection criteria}
In order to build a sample as complete as possible, we only impose two conditions that have to be met: (1) an S/N of at least 3 in the F444W band, and (2) redshifts $3 \leq z \leq 9$. Based on the nature of the redshift used (photometric or spectroscopic), we construct two samples, that we now introduce.

The photometric sample is initially composed of 37272 galaxies. We use the SED fitting code \texttt{Prospector} \citep{Johnson2019,Johnson2021} to fit the full sample (see Section~\ref{sec:Prospector}), yielding a total sample of 35442 galaxies. The cases that failed ($\sim5$\%) were due to either poor photometric coverage ($<6$ NIRCam photometric points) or to them being false detections (e.g. a diffraction spike). In order to have reliable inferred galaxy properties, we only use the results which have a reduced $\chi^2 \leq 1$, resulting in a final photometric sample with 25319 galaxies. The completeness in stellar mass and UV magnitude is discussed in Section~\ref{sec:general_properties}. We note that if we extend the limit up to $\chi^2$ = 10, we obtain 29323 galaxies instead. However, the final stellar mass complete sample is virtually unchanged from the one discussed in this work. The spectroscopic sample is composed of 1620 galaxies, all of which are also part of the photometric sample. Their redshift classification is as follows: 1234 in Category A, 363 in category B, 12 in category C, and 11 in category D.

\section{SED fitting with Prospector}
\label{sec:Prospector}
As demonstrated in \cite{Simmonds2023_JADES}, the ionising properties inferred with the SED fitting code \texttt{Prospector} \citep{Johnson2019,Johnson2021} are in good agreement with those obtained by emission line fluxes, when such fluxes are detected. Therefore, in this work, we fit the entirety of our samples with \texttt{Prospector}, without introducing an additional selection bias (i.e. by only selecting emission line galaxies).

\texttt{Prospector} uses photometry and/or spectroscopy as an input in order to infer stellar population parameters, from UV to IR wavelengths. We use photometry from the HST ACS bands: F435W, F606W, F775W, F814W, F850LP, and from the HST IR bands: F105W, F125W, F140W, and F160W. In addition, we use the JADES NIRCam photometry from: F090W, F115W, F150W, F162M, F200W, F250M, F277W, F300M, F335M, F356W, F410M, and F444W. Finally, when available, we include JEMS photometry: F182M, F210M, F430M, F460M, and F480M. The same Kron convolved aperture is used to extract the HST, JADES and JEMS photometry. 

For the photometric redshift sample, we adopt a clipped normal distribution using the \texttt{EAzY} $z_{\rm{phot}}$ as the redshift mean, with the sigma given by the $z_{\rm{phot}}$ errors. Whereas for the spectroscopic redshift sample we fix the redshift to $z_{\rm{spec}}$ for galaxies in Categories A, B, and D, and use the same prior as for the photometric sample for galaxies in Category C.

We vary the dust attenuation and stellar population properties following \cite{Tacchella2022}. In summary, we use a two component dust model described in \cite{Charlot2000} and  \cite{Conroy2009}. This model accounts for the differential effect of dust on young stars ($< 10$ Myr) and nebular emission lines, through different optical depths and a variable dust index \citep{Kriek2013}. We adopt a Chabrier \citep{Chabrier2003} initial mass function (IMF), with mass cutoffs of 0.1 and 100 M$_{\odot}$, respectively, allowing the stellar metallicity to explore a range between 0.01 - 1 Z$_{\odot}$, and include nebular emission. The continuum and emission properties of the SEDs are provided by the Flexible Stellar Population Synthesis (FSPS) code \citep{Byler2017}, based on \texttt{Cloudy} models \citep[v.13.03; ][]{Ferland2013} using MESA Isochrones \& Stellar Tracks \citep[MIST; ][]{Choi2016,Dotter2016}, and the MILES stellar library \citep{Vazdekis2015}. We note that he UV extension of the MILES library is based on the Basel Stellar Library \citep[BaSeL; ][]{Lastennet2002}. These \texttt{Cloudy} grids introduce an upper limit on the permitted ionisation parameters (log$\langle U \rangle$$_{\rm{max}} = -1.0$). We briefly remark here that this upper limit might not be appropriate for high-redshift galaxies \citep[see e.g, ][]{Cameron2023_ISM}. Due to the stochastic nature of the IGM absorption, we set a flexible IGM model based on a scaling of the Madau model \citep{Madau1995}, with the scaling left as a free parameter with a clipped normal prior ($\mu = 1.0, \sigma = 0.3$, in a range [0.0, 2.0]). Last but not least, we use a non-parametric SFH \citep[continuity SFH; ][]{Leja2019}. This model describes the SFH as eight different SFR bins, the ratios and amplitudes between them are in turn, controlled by the bursty-continuity prior \citep{Tacchella2022bursty}. For a general view of the goodness of fits of our stellar mass complete sample (described in Section~\ref{section:completeness}), we direct the reader to Appendix~\ref{app:chi}, where we show a comparison between the modelled and observed photometry, as a function of the bands being used.

Figure~\ref{fig:SED_examples} shows example SEDs and best-fit spectra for our sample. From top to bottom the redshift increases from 3.5 to 8.5, as indicated on the top left of each panel. The markers show HST (triangles) and JWST NIRCam (circles) photometry with their corresponding errors. The purple shaded area corresponds to the rest-frame spectral region at $\lambda_{\rm{rest-frame}} = 1250-2500$ \AA\/, used to estimate the rest-frame UV continuum slope \citep[$\beta$; ][]{Calzetti1994}, defined as F$_{\lambda} \propto \lambda^{\beta}$. We obtain $\beta$ by fitting a line to the best-fit SED provided by \texttt{Prospector}. We stress that there are important limitations to the measurements of $\beta$ without spectra \citep[for a full review, see ][]{Austin2024} . Finally, the vertical dashed lines show the observed wavelength of \lya\/, \oii\/, \oiii\/, and \ha\/. We note that not all galaxies have obvious emission lines detected in their photometry, illustrating the advantage of our sample selection (see for example, the second panel: JADES-GS+53.06045-27.82581).  

\section{Sample general properties}
\label{sec:general_properties}
In this section we discuss some of the general properties of our samples. We first compare the photometric redshifts inferred by \texttt{EAzY} to those obtained with \texttt{Prospector}, and to the spectroscopic redshifts (when available). We then describe the stellar mass completeness of our sample.   

\subsection{Redshift comparisons}
As previously mentioned, we use \texttt{EAzY}-inferred redshifts as priors when fitting the photometric sample. The redshifts obtained by this template-fitting tool have proven to be reliable when using the full JADES NIRCam photometry set \citep[see Figure 13 of ][]{Rieke2023}. In Figure~\ref{fig:photoz_comparisons}, we compare the \texttt{EAzY} and the \texttt{Prospector} redshifts for the photometric sample. The \texttt{Prospector} redshifts are better constrained, as seen by the median error bars (black point). This is not surprising since \texttt{Prospector} is already using the \texttt{EAzY} results as a prior on the redshift. We note that \texttt{Prospector} is more flexible than \texttt{EAzY}, since the latter uses a fixed set of templates. Although, we note that the linear combination of templates used by \texttt{EAzY} might be outside of the \texttt{Prospector} parameter space, or disfavoured by the SFH or other \texttt{Prospector} priors. It can also be seen that a lower-redshift solution is preferred for several sources. However, the distribution overall follows a one-to-one relation with a best fit slope of $0.96 \pm 0.01$ (black dashed line), demonstrating that both methods are in general agreement. We note that the number of sources decreases significantly as a function of photometric redshift: 10215 at $3 \leq z < 4$, 6005 at $4 \leq 5$, 5203 at $5 < z \leq 6$, 3037 at $6 < z  \leq 7$, 651 at $7 < z \leq 8$, and, 208 at $8< z \leq 9$.      

In order to test how well the $z_{\rm{phot}}$ retrieve real ($z_{\rm{spec}}$) redshifts, in Figure~\ref{fig:photoz_specz_comparisons} we compare the photometric redshifts (from both \texttt{EAzY} and \texttt{Prospector}) to the spectroscopic sample in Categories A and B. The spectroscopic sample is biased towards brighter galaxies with stronger emission lines, compared to the full photometric sample, so we would expect more accurate $z_{\rm{phot}}$ estimations. We find a good agreement between $z_{\rm{phot}}$ and $z_{\rm{spec}}$ (i.e.  $|z_{\rm{spec}}-z_{\rm{phot}}|<0.15$), with only a small fraction of outliers. Specifically, 44 (35) for \texttt{Prospector} (\texttt{EAzY}) derived redshifts, corresponding to $\sim 3.1$\% ($\sim 2.5$\%) of the subsample, where photometry alone makes it hard to distinguish a spectral break from another. From this comparison, we cannot say if one code performs better than the other, but we can conclude that both retrieve the correct redshift in the majority of the cases. We highlight that the SED modelling uncertainties and redshift variations are self-consistent. 

Given the agreement between $z_{\rm{phot}}$ and $z_{\rm{spec}}$, in the remainder of this work, we use the photometric sample unless explicitly stated.

\subsection{Completeness of sample}
\label{section:completeness}
To estimate the stellar mass completeness of our sample, we use the redshifts and stellar masses inferred with \texttt{Prospector}, noting that the spectroscopic sample overlaps with the photometric one, and that the redshifts are in tight agreement. A corner plot showing the mean shape of the posteriors  for these parameters can be found in Appendix~\ref{app:corner}.

To assess the 90\% stellar mass completeness limit of our photometric sample, we follow the procedure described in Section 5.2 of \cite{Pozzetti2010}. In summary, for every redshift we define a minimum mass (M$_{\rm{min}}$), above which all types of galaxies can potentially be observed. To obtain M$_{\rm{min}}$, we first need to calculate the limiting stellar mass (M$_{\rm{lim}}$) for each galaxy, given by:
\begin{equation}
    \log(\text{M}_{\rm{lim}}) = \log(\text{M}_\star)+0.4(\text{m}-\text{m}_{\rm{lim}}),
\end{equation}
where M$_{\star}$ is stellar mass in units of solar masses. M$_{\rm{lim}}$ represents the mass a galaxy would have if its apparent magnitude (m) were equal to the limiting magnitude of the survey (m$_{\rm{lim}}$) in the F444W band. This band was chosen since it has the longest effective wavelength, and thus, is the tracer of stellar mass. We divide our data into three depths depending on the exposure time: medium (\texp\/ < 25 ks), deep (25 ks $\leq$ \texp\/ < 65 ks), and ultra-deep (\texp\/ > 65 ks), with 5$\sigma$ flux depths of 6 nJy, 4.5 nJy, and 2.65 nJy, respectively. Once M$_{\rm{lim}}$ has been calculated for every galaxy, we compute M$_{\rm{min}}$ for each field by selecting the faintest 20\% of the galaxies with a given \texp\/, and finding the limit above which 90\% of the selected M$_{\rm{lim}}$ values lie. Figure~\ref{fig:completeness} shows the stellar mass of our sample as a function of redshift (grey circles), the coloured small circles show the M$_{\rm{lim}}$ of the faintest 20\% of galaxies for each field, and the filled lines denote the 90\% completeness as a function of redshift. We find that, in general, our photometric sample is stellar mass complete down to log(M$_{\star}$/[M$_{\odot}$])$\approx$7.5, with slight variations depending on the depth of the observations (within $\sim 0.2$ dex). 

In the remainder of this study, we focus only on sources that have a \texttt{Prospector}-derived stellar mass with a median above the mass completeness limit (log(M$_{\star}$/[M$_{\odot}$])$\approx$7.5), yielding a sample of 14652 galaxies, which are UV complete down to M$_{\rm{UV}}\approx-16$.

\subsection{M$_{\rm{UV}}$-M$_{\star}$ relation}

Figure~\ref{fig:MUV_mass_zbins} shows the stellar mass of our stellar mass complete sample as a function of M$_{\rm{UV}}$, divided by redshift bins. The numbers on the top of each panel indicate the number of galaxies in any given bin. We fit the observational data in the M$_{\rm{UV}}$-M$_{\star}$ plane taking into account the completeness limits discussed in the previous section. Specifically, we fit the following relation
\begin{equation}
    \log(\text{M}_{\star}) = \alpha (\text{M}_{\rm UV} + 19.5) + \log(\text{M}_{\star,0}),
\label{eq:MuvMs}
\end{equation}
where M$_{\star}$ is stellar mass in units of solar masses, $\alpha$ is the slope of the M$_{\rm{UV}}$-M$_{\star}$ relation and $\log(M_{\star,0})$ is its normalisation (stellar mass at $M_{\rm{UV}}=-19.5$ AB mag). In addition to these two parameters, we also fit for the scatter $\sigma_{\rm M_{\rm{UV}}-M_{\star}}$ in M$_{\rm{UV}}$ at fixed M$_{\star}$. We then fit these three parameters to the observed distribution using the dynamic nested sampling code \texttt{dynesty} \citep{Speagle2020}. In each model call, we sample the M$_{\rm{UV}}$-M$_{\star}$ relation (with scatter) with 1 million galaxies assuming an M$_{\rm{UV}}$ that follows the UV luminosity function \citep{Bouwens2021}, bin this drawn galaxy sample in the M$_{\rm{UV}}$-M$_{\star}$ plane taking into account the completeness limits, and compare the normalised histogram with the observed one. The best-fit parameters can be found in Table~\ref{tab:bestfit_M_MUV}, for the highest redshift bin ($8 < z \leq 9$) we adopt the best-fit results from the previous redshift bin ($7 < z \leq 8$).

From the nearly linear relation between $\log(\text{M}_{\star})$ and $\log(\text{SFR})$ \citep[i.e.  star-forming main sequence; ][]{Brinchmann2004,Daddi2007,Salim2007}, we would expect a slope $\alpha$ of the M$_{\rm{UV}}$-M$_{\star}$ relation close to $-0.4$, assuming a simple, linear conversion between the UV luminosity and SFR \citep{Kennicutt1998,Madau1998,Salim2007}. Our steeper slope of $\approx-0.6$ is consistent with having a higher mass-to-light ratio ($M_{\star}/L_{\rm UV}$) for more massive systems, in agreement with the findings of \cite{Gonzalez2011}.
We further find that $\log(M_{\star,0})$ decreases with increasing redshift ($M_{\star,0}\propto(1+z)^{-2.7}$), i.e. at fixed M$_{\rm UV}=-19.5$ the typical stellar mass of the galaxies is lower at earlier cosmic times. This implies a lower $M_{\star}/L_{\rm UV}$ toward higher redshifts, consistent with younger stellar populations toward earlier times. A combination of less dust attenuation and higher specific SFR toward higher redshifts and at fixed stellar mass could lead to this outcome. Finally, the scatter of the M$_{\rm{UV}}$-M$_{\star}$ relation is substantial with $0.6-0.7$ dex. This scatter includes both the intrinsic scatter of the M$_{\rm{UV}}$-M$_{\star}$ relation and the observational uncertainty, though the latter is only of the order of 0.2 dex. 

\begin{table}
    \caption{Best-fit parameters for log(M$_{\star}$)-M$_{\rm{UV}}$ relation as parameterised by Eq.~\ref{eq:MuvMs}. \textsl{Column~1:} redshift bin. \textsl{Column~2:} slope of the relation. \textsl{Column~3:} stellar mass normalisation (log(M$_{\star}$) at M$_{\rm UV}=-19.5$. \textsl{Column~4:} scatter in log(M$_{\star}$) at fixed M$_{\rm UV}$.}
    \centering
    \begin{tabular}{c|c|c|c}
    \hline
    \noalign{\smallskip}
    Redshift & $\alpha$ & log(M$_{\star,0}$/[M$_{\odot}$]) & $\sigma$  \\
     & & & [dex] \\
    \noalign{\smallskip}
    \hline
    \noalign{\smallskip}
         $3 \leq z \leq 4$ & $-0.60^{+0.27}_{-0.33}$ & $9.25^{+0.77}_{-0.71}$ & $0.66^{+0.31}_{-0.28}$\\
         $4 < z \leq 5$ & $-0.69^{+0.26}_{-0.28}$ & $9.06^{+0.62}_{-0.63}$ & $0.69^{+0.29}_{-0.28}$\\
         $5 < z \leq 6$ & $-0.55^{+0.20}_{-0.32}$ & $8.78^{+0.71}_{-0.60}$ & $0.57^{+0.37}_{-0.22}$ \\
         $6 < z \leq 7$ & $-0.65^{+0.21}_{-0.30}$ & $8.63^{+0.51}_{-0.46}$ & $0.56^{+0.37}_{-0.21}$ \\
         $7 < z \leq 8$ & $-0.68^{+0.25}_{-0.28}$ & $8.55^{+0.57}_{-0.57}$ & $0.72^{+0.26}_{-0.26}$ \\
         \hline
    \end{tabular}
    \label{tab:bestfit_M_MUV}
\end{table}

As comparison, we show the log(M$_{\star}$)-M$_{\rm{UV}}$  mass-to-light relations from \cite{Duncan2014}, \cite{Song2016}, and \cite{Tacchella2018}. The \cite{Duncan2014} and \cite{Song2016} relations are derived from a combination of observations (i.e. Spitzer/IRAC and HST) and SED fitting, while the one from \cite{Tacchella2018} is based on an empirical model that links star formation in galaxies to the accretion rate of dark matter halos. With our stellar mass complete sample, we find a somewhat steeper slope than previous observational studies (but mostly $<1\sigma$ discrepant; $\sim-0.6$ instead of $\sim -0.4$ to $-0.5$). This effect is partly a consequence of our sample including a larger population of fainter sources (with lower stellar masses) than the mentioned studies, but also because we consider explicitly the mass completeness of our sample. Specifically, we forward model the completeness when fitting the relation; if we do not do that, we find a shallower relation of $\sim -0.4$. Additionally, the stellar masses in these previous studies were estimated assuming parametric SFHs, which cannot fully describe the complexities of galaxies \citep{Lower2020}, while we assume more flexible SFHs.

\begin{table*}
    \caption{Table excerpt of general properties for a selection of galaxies studied in this work. \textsl{Column 1:} JADES identifier, composed of the coordinates of the centroid rounded to the fifth decimal place, in units of degrees. \textsl{Column 2:} photometric redshift inferred using the SED fitting code \texttt{Prospector}. \textsl{Column 3:} logarithm of stellar mass. \textsl{Column 4:} logarithm of the ionising photon production efficiency. \textsl{Column 5:} logarithm of the rate of ionising photons being emitted. \textsl{Column 6:} logarithm of the burstiness of SFH, defined as the ratio between recent star formation (<10 Myr) and the one averaged over the past 100 Myr (using the median values of SFR$_{10}$ and SFR$_{100}$). \textsl{Column 7:} observed UV magnitude.}
    \centering
    \begin{tabular}{ccccccc}
    \hline
    \noalign{\smallskip}
    Name & $z$ & log(M)& log(\xionnofesc\/) & log(\ndot\/) & log(SFR$_{10}$/SFR$_{100}$)& M$_{\rm{UV,obs}}$ \\
        &    & [M$_{\odot}$] & [Hz erg$^{-1}$]  & [s$^{-1}$] & & [AB] \\
    \noalign{\smallskip}
    \hline
    \noalign{\smallskip}
    \input{Table1_excerpt.dat}
    \end{tabular} 
    \label{table:properties}
\end{table*}

\section{Constraints on the ionising properties of galaxies}
\label{sec:ionising_properties}
In this section, we provide a brief overview of the methods traditionally used to infer ionising properties of galaxies. Followed by the trends of \xion\/ and \ndot\/ with redshift and UV magnitude, including a comparison with values from the literature. Finally, we discuss the emergence of a secondary (previously unseen) sample of galaxies, that are tentatively leaking LyC. An excerpt of the properties used in this section is given in Table~\ref{table:properties}.

\subsection{Measuring \xion\/ from emission line fluxes}
The ionising photon production efficiency, \xion\/, can be measured through \ha\/ and/or \oiii\/ emission line fluxes. In order to use \ha\/ as a proxy for ionising photon production efficiency (\xion\/), one must first correct the \ha\/ line flux for dust \citep[for example, using an SMC attenuation curve; ][]{Gordon2003}. If Case B recombination is assumed, along with the assumption of no LyC \fesc\/, and that dust has a negligible effect on LyC photons, the dust corrected \ha\/ luminosity relates directly to the rate of ionising photons (\ndot\/, in units of s$^{-1}$) that are being emitted \citep[\ndot\/ $= 7.28 \times 10^{11}$ L(\ha\/); ][]{Osterbrock2006}. \xion\/ is then the ratio between \ndot\/ and the observed monochromatic UV luminosity density (measured at rest-frame $\lambda = 1500$ \AA\/). This method suffers from a number of of assumptions that need to be made: it has a great dependence on the chosen attenuation curve, and Case B recombination cannot always be assumed \citep[][]{McClymont2024,Scarlata2024}. If \oiii\/$_{\lambda 5007}$ is available instead, its equivalent width can give a measure of the ionisation field of the galaxy as shown in \cite{Chevallard2018}, and later in \cite{Tang2019}. This method also depends on the shape of the attenuation curve assumed, as well as on the gas-phase metallicity and nebular conditions.

In \cite{Simmonds2023_JADES}, the ionising properties of a sample of 677 emission line galaxies at $z\sim4-9$ were analysed. These galaxies had signatures of either \ha\/ and/or \oiii\/ emission in their NIRCam photometry, specifically with a 5$\sigma$ detection in the filter pairs F335M-F356W and F410M-F444W. Both methods mentioned above were used to estimate \xion\/, while in parallel, \texttt{Prospector} was used to fit their full NIRCam photometry (assuming \fesc\/ = 0). A tight agreement was found between the values measured by emission line fluxes and those inferred by  \texttt{Prospector}. Therefore, in this work we rely on this SED fitting code to extract the ionising properties of a stellar mass complete sample, allowing us to potentially study all types of galaxies, not only those with detectable emission lines. Since our \texttt{Prospector} fitting routine assumes an escape fraction of zero, our estimated ionising photon production efficiencies (\xionnofesc\/) include a correcting factor of $1-$\fesc\/, such that:
\begin{equation}
\xi_{\rm{ion,0}} = \xi_{\rm{ion}}\times (1-\text{f}_{\rm{esc}}).
\end{equation}
Appendix~\ref{app:IG} shows how well \texttt{Prospector} can constrain stellar masses and ionising photon production efficiencies, as a function of flux in F444W.

\subsection{Trends of ionising properties with redshift}
We first compare our results to those from the literature, in particular, for the evolution of \xionnofesc\/ with redshift. Figure~\ref{fig:xion_literature} encodes most results to date. \cite{Stefanon2022} compiled \xionnofesc\ measurements up to $z\sim8$ \citep[containing data from ][]{Stark2015,Stark2017,Marmol-Queralto2016,Nakajima2016,Bouwens2016,Matthee2017,Harikane2018,Shivaei2018,DeBarros2019,Lam2019,Faisst2019,Tang2019,Nanayakkara2020,Emami2020,Endsley2021,Atek2022,Naidu2022}. This extensive compilation allowed them to estimate the rate of change of \xionnofesc\/ with redshift, finding it to have a slope of dlog(\xionnofesc\/)/d$z=0.09\pm 0.01$ (black dashed line). 
The markers show the UV-faint galaxies at $z\sim 3-7$ from \cite{Prieto-Lyon2023}, the Lyman-$\alpha$ emitters (LAEs) from \cite{Ning2023} and \cite{Simmonds2023_JEMS}, the \ha\/ emitters from \cite{Rinaldi2024HAEs}, and the ELGs from \cite{Simmonds2023_JADES}. The latter show a slightly less steep evolution of \xionnofesc\/ with redshift than the compilation of \cite{Stefanon2022}, but consistent within errors (given by dlog(\xionnofesc\/)/dz$=0.07\pm 0.02$). Finally, the stellar mass complete sample of this work is shown as grey pentagons. There is an overlap with previous measurements, however, there is a population of previously unseen galaxies with log(\xion\//[Hz erg$^{-1}$]) $\lesssim 24.5$. The inclusion of these galaxies has the result of flattening of the increase of \xionnofesc\/ with redshift. 

Figure~\ref{fig:xion_nion_redshift} shows \xionnofesc\/ and \ndot\/ as a function of redshift, for our photometric (no edges) and spectroscopic (black edges) sample, colour-coded by the burstiness of their SFH \citep[which correlates the most with \xionnofesc\/; ][]{Simmonds2023_JADES}. The crosses represent the galaxies with no recent star formation (SFR$_{10} = 0$ M$_{\odot}$ yr$^{-1}$). As a reminder, both samples overlap and the \texttt{Prospector}-inferred redshifts agree with the $z_{\rm{spec}}$ ones, for a great majority of the overlapping galaxies. We present both samples fit separately to highlight that the increase in \xionnofesc\/ observed previously is mostly due to a selection effect (by selecting predominantly star-forming galaxies with emission lines). Indeed, the best fit to \xionnofesc\/ for the spectroscopic sample has a slope of dlog(\xionnofesc\/)/dz$\sim0.04\pm 0.02$, within errors of the findings of \cite{Simmonds2023_JADES}. Importantly, when the full photometric sample is taken into account, the slope is considerably flatter (dlog(\xionnofesc\/)/dz$\sim-0.02\pm 0.00$). The same effect can be seen in the best-fit lines to \ndot\/. 

\subsection{Trends of ionising properties with UV magnitude}
\xion\/ depends on several galaxy properties, such as metallicity, age, and dust content \citep{Shivaei2018}. Moreover, fainter galaxies have been shown to be more efficient in producing ionising radiation \citep[e.g. ][]{Duncan2015,Maseda2020,Simmonds2023_JADES,Endsley2024}. Figure~\ref{fig:xion_MUV_zbins} shows \xionnofesc\/ as a function of M$_{\rm{UV}}$ for our data, per redshift bin. We first note that as redshift increases, there are less galaxies per bin, and they tend to be fainter. We find that by dividing the sample into "star-forming" (log(SFR$_{10}$/SFR$_{100}$) $\geq$ -1) and "mini-quenched"
(log(SFR$_{10}$/SFR$_{100}$) < -1), we distinguish two populations of galaxies that have consistent \xionnofesc\/ slopes with M$_{\rm{UV}}$, but populate a different \xionnofesc\/ range. The star-forming sample lies above log(\xionnofesc\//[Hz erg$^{-1}$])$\sim 24.5$, while the mini-quenched \citep[see e.g. ][]{Looser2023a,Looser2023b} sample mostly lies below. The number of galaxies in each bin is shown in the top of the panels, with format \textsl{N = XX (YY)}, where \textsl{XX} is the number of star-forming galaxies and \textsl{YY} is the number of mini-quenched galaxies. The best fit to both datasets (when enough points to fit a line reliably) are shown in the bottom of each panel. We find that in general, there is a slight increase of \xionnofesc\/ towards the fainter galaxies (slope of $\sim 0.01-0.04$), but this increase is lower than the slope of $\sim 0.1$ found in \cite{Simmonds2023_JADES}. This is unsurprising given the nature of our sample (i.e. containing potentially all types of galaxies). When taking these two populations in mind, the \xionnofesc\/-$z$ relation shown in the top panel of Figure~\ref{fig:xion_nion_redshift} (for the photometric sample) becomes:
\begin{equation}
\label{eq:xion_z}
    \log (\text{\xionnofesc\/(z)}) = (-0.001\pm 0.004)z + (25.294\pm 0.017)
\end{equation}
for the star-forming sample. Likewise, the \ndot\/-$z$ relation shown in the bottom panel of Figure~\ref{fig:xion_nion_redshift}, for the star-forming sample becomes:
\begin{equation}
    \log (\text{\ndot\/(z)}) = (0.032\pm 0.016)z + (53.106\pm 0.079).
\end{equation}

As with \xionnofesc\/, the mini-quenched galaxies also appear to populate a distinct region in the \ndot\/ versus UV magnitude plane. Contrary to \xionnofesc\/, however, \ndot\/ has a steeper dependence with M$_{\rm{UV}}$ (with a slope of $\sim -0.4$), as shown in Figure~\ref{fig:nion_MUV_zbins} \footnote{Appendix~\ref{app:tables} contains the relations shown in Figures~\ref{fig:xion_MUV_zbins} and~\ref{fig:nion_MUV_zbins} in table format.}. In summary, as galaxies become fainter, \xionnofesc\/ marginally increases, but \ndot\/ significantly decreases. This is an effect that derives naturally from the definition of \xionnofesc\/: the ratio between the ionising photons being produced (\ndot\/) and the non-ionising UV continuum luminosity. The former is dominated by the contribution of young hot stars, while the latter also includes the contribution of older stellar populations. 

Finally, we combine our redshift and UV magnitude relations to perform a 2-dimensional fit and find a joint relation for the star-forming sample, given by:
\begin{multline*}
\log(\xi_{\rm{ion}} (z,\text{M}_{\rm{UV}})) = \\ 
    (0.003\pm 0.003)z + (-0.018\pm 0.003)\text{M}_{\rm{UV}} + (25.984\pm 0.053)
\end{multline*}

\subsection{Unveiling the silent population}
We find a total of 350 mini-quenched galaxy candidates, corresponding to $\sim 2.4$\% of our total sample. Their number increases with redshift up to $z = 7$ and then decreases significantly. There are a few possible explanations for this behaviour that most likely are working together. On one hand, the preference of mini-quenched galaxies in the redshift window of $z=4-6$ could have a physical origin \citep{Dome2024}: at higher redshifts, galaxies are bursty and undergo quenching attempts, but the replenishment time of gas within galaxies is so short that the mini-quenching phase is not observable (i.e. less than a few Myr). Supporting this scenario, recent work by \cite{Witten2024} shows a $z\sim 7.9$ galaxy with a bursty SFH, with evidence of a mini-quenched episode lasting $\sim 20$ Myr followed by a rejuvenation event. Towards lower redshifts, the mini-quenching phase might have a longer duration or SFHs are less bursty, leading to fewer mini-quenched systems.  
On the other hand, at $z \sim 5-7$, the UV part of the spectrum is shifted into the NIRCam F090W filter, which covers our entire stellar mass complete sample (see Figure~\ref{fig:filter_curves}). The deep NIRCam observations allow for stronger constraints on the SED shape, and could explain the increased detection of mini-quenched galaxies at these redshifts. Finally, at higher redshifts ($z > 7$), it becomes increasingly difficult to detect these almost featureless galaxies with photometry alone (and with a S/N > 3).

"Mini-quenched galaxies" have low \xionnofesc\/, well below the observed \xionnofesc\/ relations (i.e. with redshift and M$_{\rm{UV}}$), and are likely only minor players in the reionisation of the Universe. Their photometry show little-to-no evidence of emission lines, which explains why they have been neglected in previous studies of this nature. These galaxies are possibly analogues to the galaxies studied in \cite{Looser2023a,Strait2023} and \cite{Looser2023b}, which are temporarily quenched due to their extremely bursty SFHs \citep{Dome2024}. Interestingly, using simulations, \citet{Dome2024} predict the number of mini-quenched galaxies increases with cosmic time, reaching $\sim2-4$\% at $z=4$, in broad agreement with our preliminary findings. A spectroscopic follow-up study is needed to confirm this hypothesis. 

\begin{figure}
    \includegraphics[width=\columnwidth]{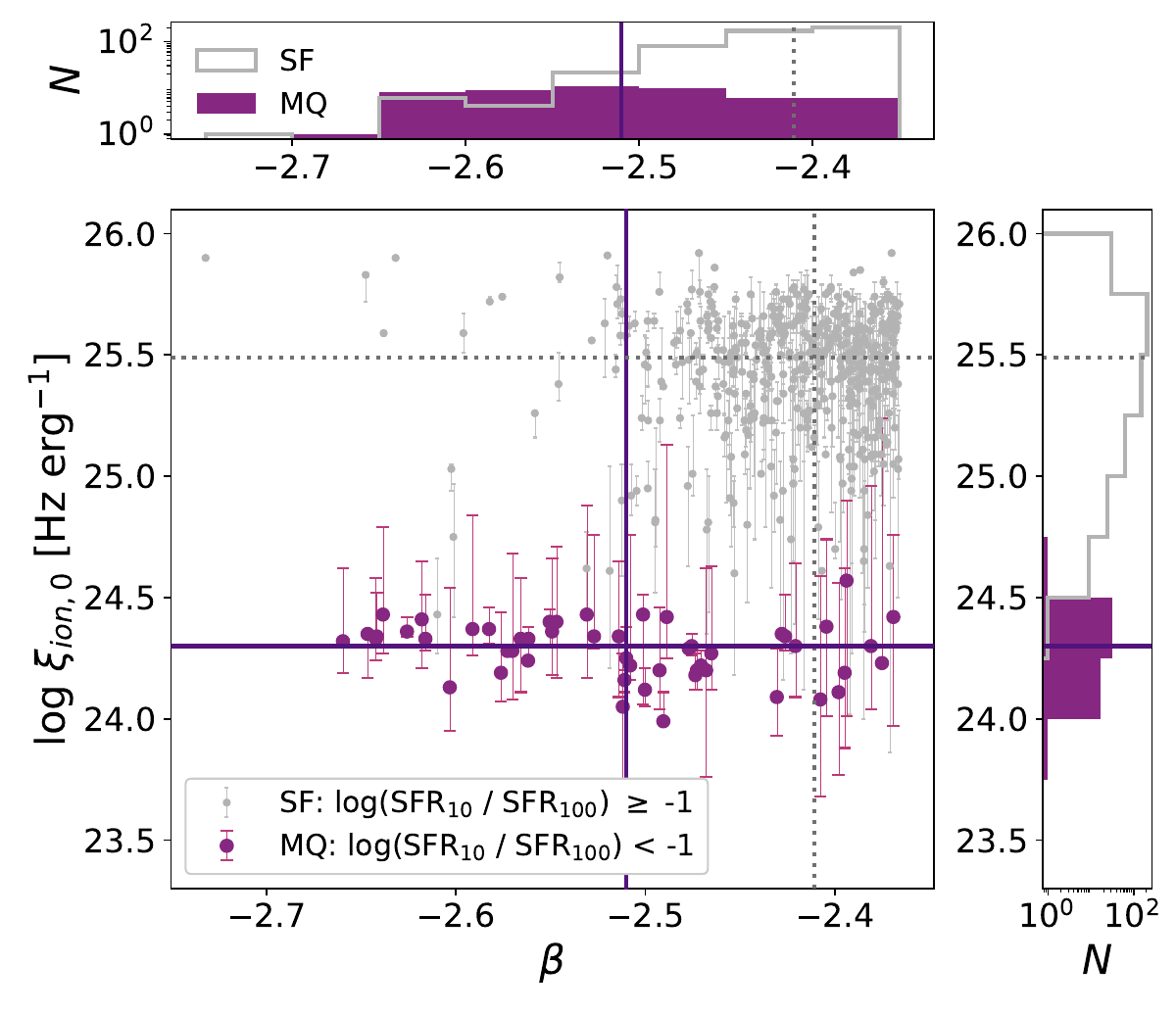}
    \caption{\xionnofesc\/ versus $\beta$ for the galaxies with \fesc\/(LyC) > 10\%, adopting the relation from \citet{Chisholm2022}. The two colours represent star-forming ("SF"; smaller grey circles) and mini-quenched ("MQ": purple larger circles) populations, as shown in the legend. The \xionnofesc\/ and $\beta$ medians of each sub-sample are shown as filled (mini-quenched) and dotted (star-forming) lines. Although, by definition, every galaxy with \fesc\/(LyC) > 10\% has a blue UV continuum slope, the mini-quenched candidates have on average bluer $\beta$ and lower \xionnofesc\/ values than the star-forming galaxies.}
    \label{fig:xion_beta_remnants}
\end{figure}

If we adopt the $\beta$-\fesc\/(LyC) relation from \cite{Chisholm2022}, given by:
\begin{equation}
\label{eq:fesc_Chisholm}
    \text{f}_{\rm{esc}}\text{(LyC)} = (1.3\pm 0.6) \times 10^{-4} \times 10^{(-1.22\pm 0.1)\beta},
\end{equation}
we find that a significant number of the mini-quenched galaxy candidates [51 (141)] indicate leakage of ionising radiation [\fesc\/(LyC) > 10\% (5\%)]. Corresponding to $\sim 16$\% (40\%) of the mini-quenched sample. By comparison, the star-forming sample only has $\sim$4\% (32\%) of galaxies that obey the same criteria. Figure~\ref{fig:xion_beta_remnants} shows \xionnofesc\/ as a function of $\beta$ for all galaxies with \fesc\/(LyC) > 10\%. The mini-quenched candidates populate a different parameter space in \xionnofesc\/ and $\beta$ than the star-forming galaxies, with the mini-quenched galaxies having on average bluer UV continuum slopes and lower \xionnofesc\/ values. We now focus on the strongest leaker candidates within the mini-quenched sample (\fesc\/ > 10\%).

Observing LyC leakage directly is impossible at high redshift, the average IGM transmission of hydrogen ionising photons emitted at $\lambda_{\rm{rest-frame}} \sim 900$ \AA\/, at $z \sim 6$, is virtually zero \citep{Inoue2014}. It is even more complicated to observe hydrogen ionising photons emitted at shorter wavelengths, for example at $\sim 700$ \AA\/, where the nebular contribution does not contaminate \fesc\/ estimations \citep{Simmonds2024LyC}. Therefore, indirect methods are required to understand how ionising photons escape in the early Universe. In order to confirm if these galaxies are indeed leaking, we run \texttt{Prospector} on this subsample, with a modified approach that includes LyC leakage in the fitting routing (Stoffers et al. in prep). This is work in progress and is in the process of being calibrated. Promisingly, \texttt{Prospector} finds signs of leakage for these LyC leaking candidates (finding \fesc\/ > 20\% for all of them).

\begin{figure}
    \includegraphics[width=\columnwidth]{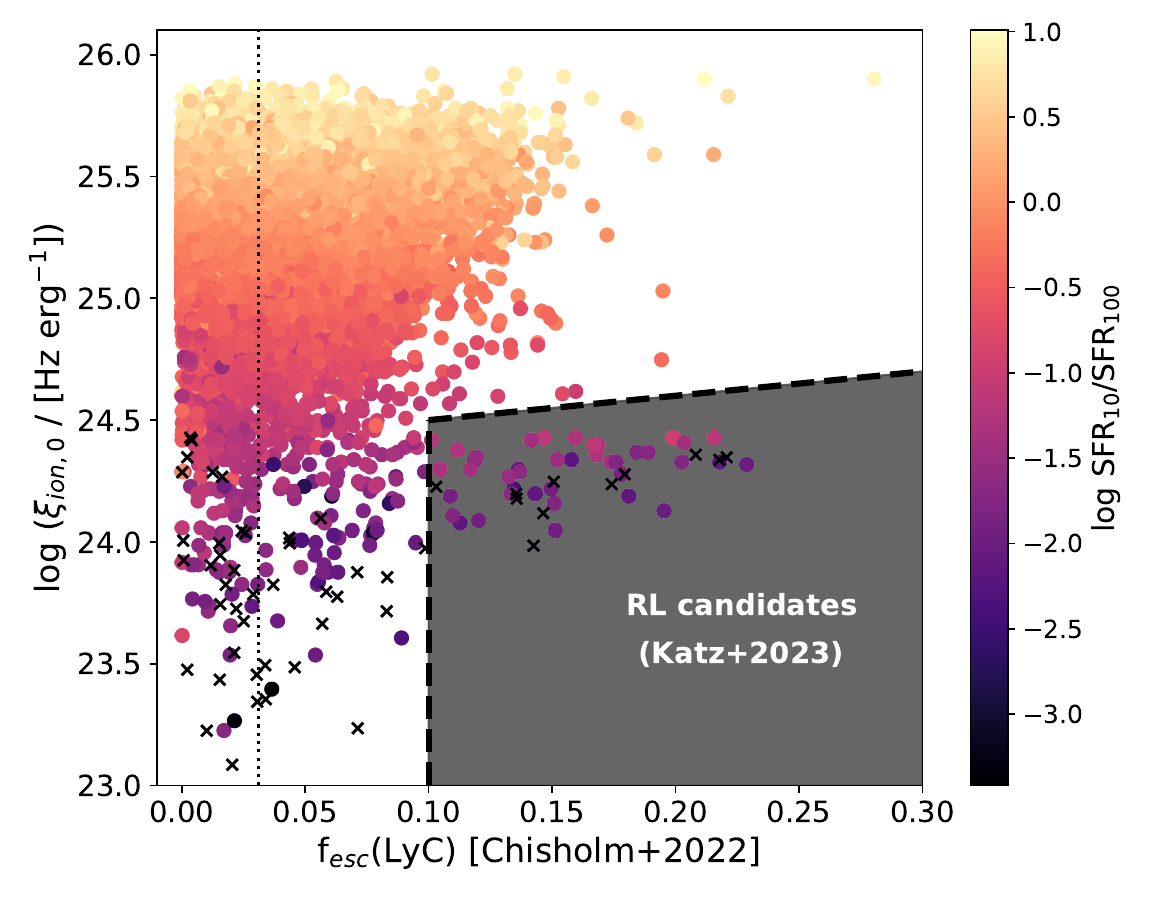}
    \caption{\xionnofesc\/ versus LyC escape fractions, estimated using the $\beta$ relation from \citet{Chisholm2022}, and colour-coded by burstiness. The vertical dashed line shows the mean \fesc\/ of the star-forming sample. As in previous figures, the crosses show the galaxies with zero recent star formation ($\text{SFR}_{10} = 0\ \text{M}_{\odot}\ \text{yr}^{-1}$). The dark shaded area highlights the region where log(SFR$_{10}$/SFR$_{100}$) < -1, and f$_{\rm{esc}}$(LyC) > 10\%, which broadly coincides with the region where log(\xionnofesc\//[Hz erg$^{-1}$]) is below 24.5. This parameter space is analogue to the one populated by the "remnant leakers" (RL) presented in \citet{Katz2023}. Following this criteria, we find 51 remnant leaker candidates.}
    \label{fig:xion_fesc_remnants}
\end{figure}

Figure~\ref{fig:xion_fesc_remnants} shows \xionnofesc\/ versus \fesc\/(LyC), colour-coded by burstiness. The shaded region shows the area where \fesc\/ > 10\%, and log(SFR$_{10}$/SFR$_{100}$) $< -1$ (the latter roughly coincides with log(\xionnofesc\//[Hz erg$^{-1}$]) < 24.5). This parameter space is analogous to the one presented in \cite{Katz2023} for "remnant leakers". In short, \cite{Katz2023} propose two modes of LyC leakage (\fesc\/ > 20\%), based on galaxies from the SPHINX suite of cosmological simulations \citep{Rosdahl2018,Rosdahl2022}. "Bursty leakers" are galaxies with a recent burst of star formation (within the last 10 Myr, SFR$_{10}$>SFR$_{100}$), akin to an ionisation bounded HII region with holes. While "remnant leakers" had a strong burst of star formation in the past, but not recently (SFR$_{100}$>SFR$_{10}$). Making remnant leakers similar to density bounded HII regions, where the ISM was disrupted enough to halt star formation. 

\begin{figure}
    \includegraphics[width=\columnwidth]{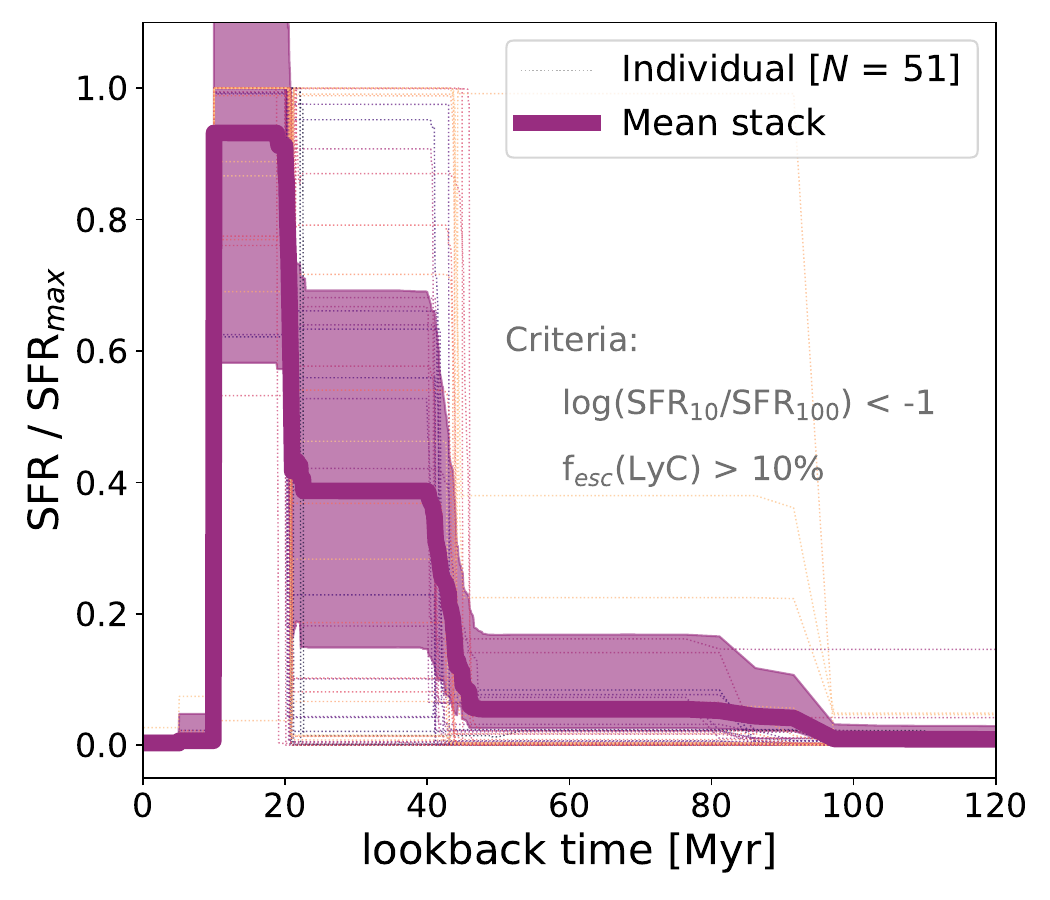}
    \caption{Stacked SFHs of the 51 galaxies populating the "remnant leakers" parameter space shown in Figure~\ref{fig:xion_fesc_remnants}. The SFHs have been normalised to their maximum, the individual SFHs are shown as thin coloured curves, and the stack as a purple thick curve. The shape of the mean SFH of these galaxies indicates a burst of star formation in their recent past (<50 Myr), but not within their immediate past (<10 Myr).}
    \label{fig:remnant_candidates}
\end{figure}

Figure~\ref{fig:remnant_candidates} shows the stacked SFHs  of the 51 remnant leaker candidates found in this work, which indicates the presence of a burst in star formation occurring in the past (within the last $\sim 50$ Myr), but no recent star formation (within the last 10 Myr), in agreement with the SPHINX remnant leakers from \cite{Katz2023}. As expected, the photometry for these candidates suggests little-to-no presence of emission lines, which could support a high \fesc\/ scenario \citep{Zackrisson2017}. The evidence of these galaxies being in a remnant leaker mode is compelling. An in depth study of these sources will be performed in the future and is beyond the scope of this work. However, given the small amount of galaxies we find to fall into the mini-quenched category (< 3\% of the total sample), and their low \xionnofesc\/, we conclude that they are not dominant agents in the reionisation of the Universe.

\begin{figure*}
    \includegraphics[width=2\columnwidth]{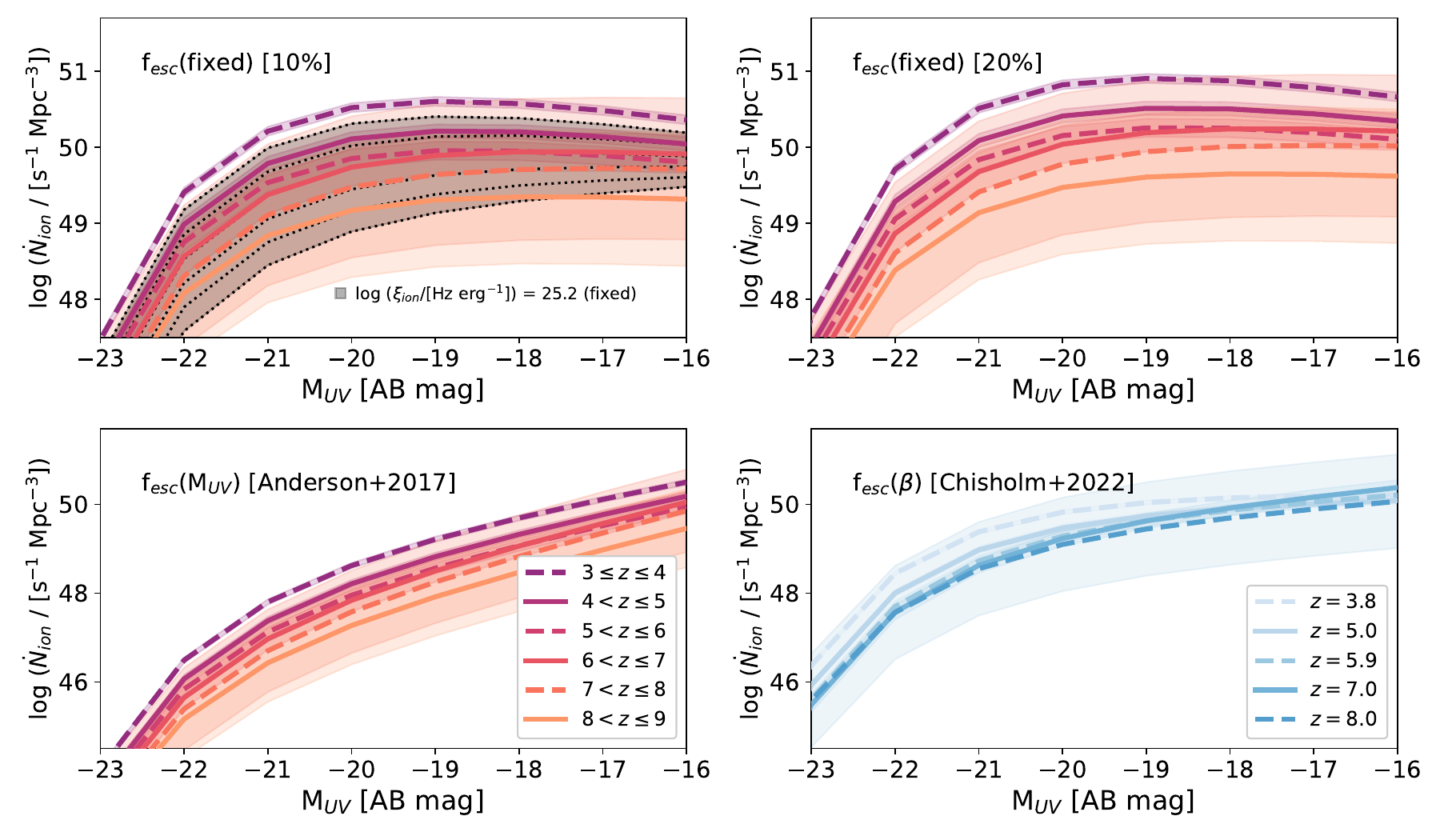}
    \caption{\Ndot\/ as a function of M$_{\rm{UV}}$, separated by redshift bins and assuming a \fesc\/ as indicated in each panel. The \xionnofesc\/ relations with M$_{\rm{UV}}$ shown in Figure~\ref{fig:xion_MUV_zbins} are adopted (using the relations given for galaxies with log(SFR$_{10}$/SFR$_{100}$) $\geq$ -1, which are representative of the general population). The variable \fesc\/ prescriptions come from \citet{Anderson2017} and \citet{Chisholm2022}, and the luminosity functions are taken from \citet{Bouwens2021}. In the top panel, we show for comparison the results adopting a fixed log(\xionnofesc\//[Hz erg$^{-1}$]) = 25.2 (grey shaded area and dotted lines).}
    \label{fig:LF_MUV}
\end{figure*}

\begin{figure*}
    \includegraphics[width=2\columnwidth]{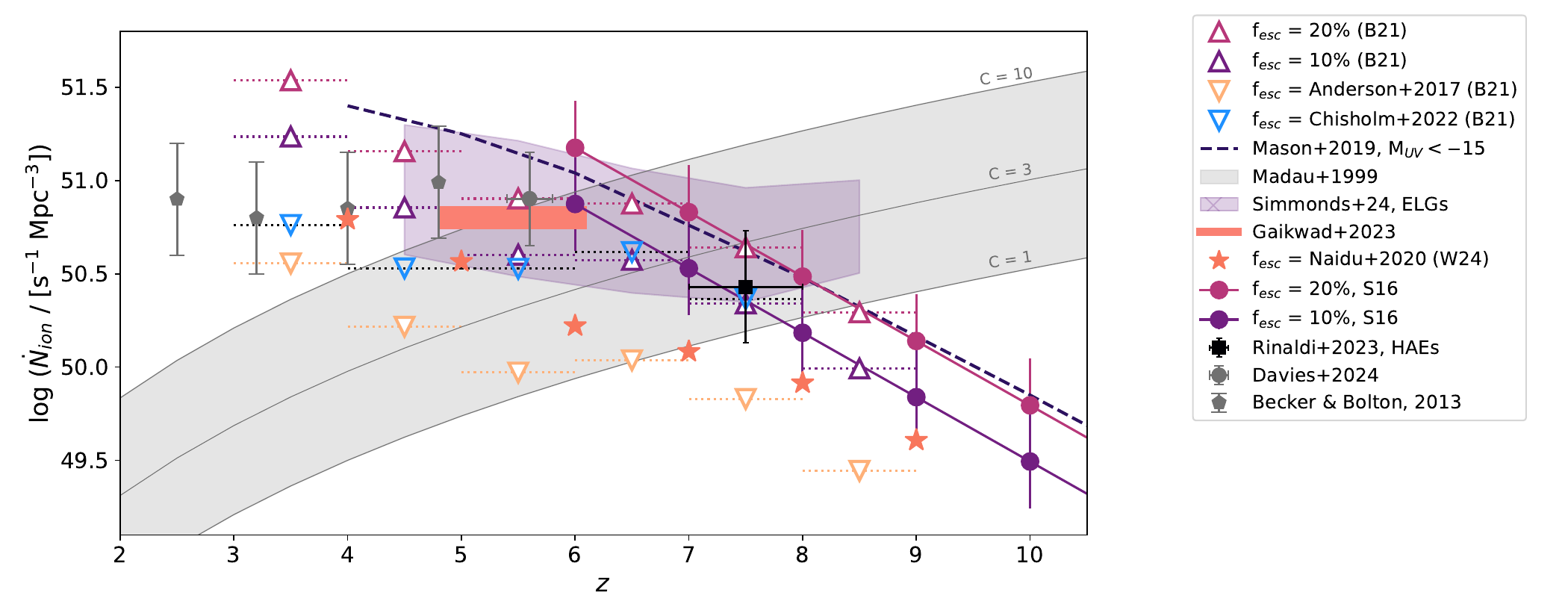}
    \caption{Cosmic rate of ionising photons emitted per second and per unit volume (\Ndot\/), as a function of redshift. The error bars in our estimations have been omitted for readability, and are on average $\sim 0.3$ dex. The same \fesc\/ prescriptions assumed in Figure~\ref{fig:LF_MUV} are adopted, as indicated in the legend. The results obtained by adopting the UV luminosity density from \citet{Sun2016} are shown as filled circles ("S16"), and those obtained by integrating the UV luminosity density curves from \citet{Bouwens2021} down to M$_{\rm{UV}} = -16$ are shown as triangles ("B21", curves shown in Figure~\ref{fig:LF_MUV}). The stars have been obtained by convolving the stellar mass functions from \citet{Weibel2024} instead ("W24"). We include the curve from \citet{Mason2015} assuming constant \xion\ and \fesc\/, integrated down to a M$_{\rm{UV}}$ of -15 \citep[as in ][]{Mason2019}, as well as the \Ndot\ reported in \citet{Rinaldi2024HAEs} for \ha\ emitters at $z \sim 7-8$ (black square). We also include the estimated \Ndot\ needed to maintain Hydrogen ionisation in the IGM \citep{Madau1999}, adopting clumping factors of 1, 3 and 10. Finally, as comparison, we show the results from \citet{Simmonds2023_JADES}, as a shaded hatched area. These were calculated under the assumption that the ELGs studied in that work were representative of the general galaxy population, with low faint low-mass bursty galaxies creating a turnover in \Ndot\/ at $z > 8$. Finally, we include observational constraints obtained by observing the \lya\/ forest from \citet{Becker_Bolton2013}, \citet{Gaikwad2023}, and \citet{Davies2024}. With our stellar mass complete sample, we find that galaxies produce enough ionising radiation to ionise the Universe by $z \approx 5-6$, without producing an excess in the cosmic ionising photon budget. Importantly, the points estimated adopting the \citet{Chisholm2022} \fesc\/ prescription, flatten at $z \lesssim 6$, in general agreement with the \lya\/ forest constraints.}
    \label{fig:cosmic_Ndot}
\end{figure*}

\begin{figure*}
    \includegraphics[width=2\columnwidth,trim={0 0 0 0}]{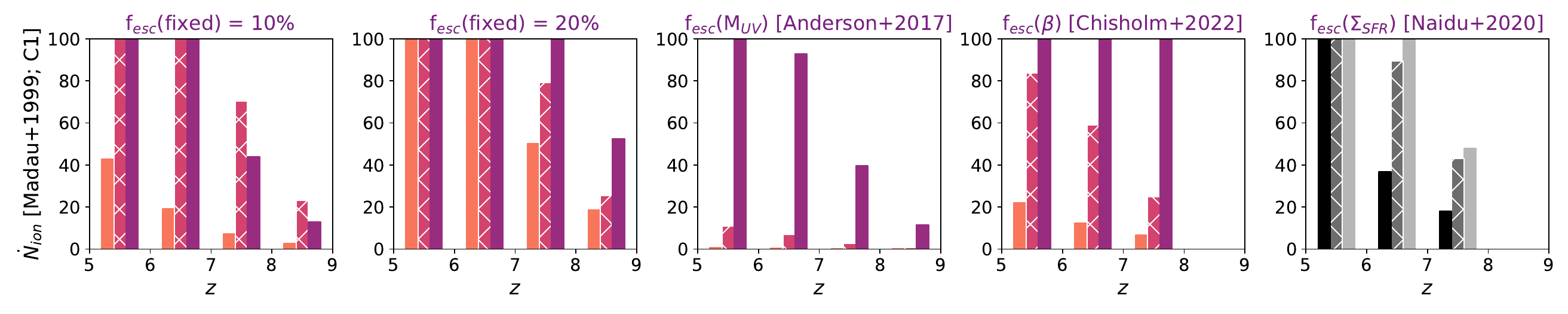}
    \includegraphics[width=2\columnwidth,trim={0 0 0 1.1cm}]{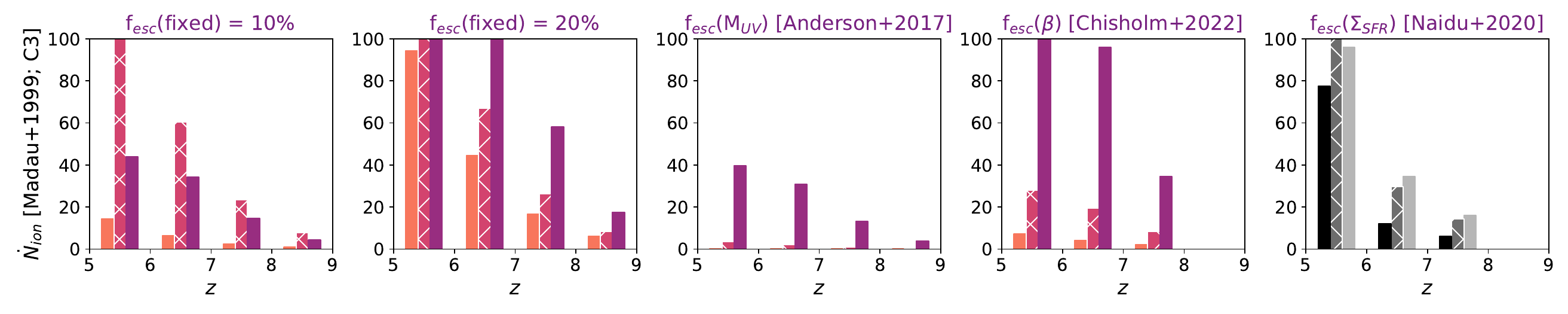}
    \includegraphics[width=2\columnwidth,trim={0 0 0 1.1cm}]{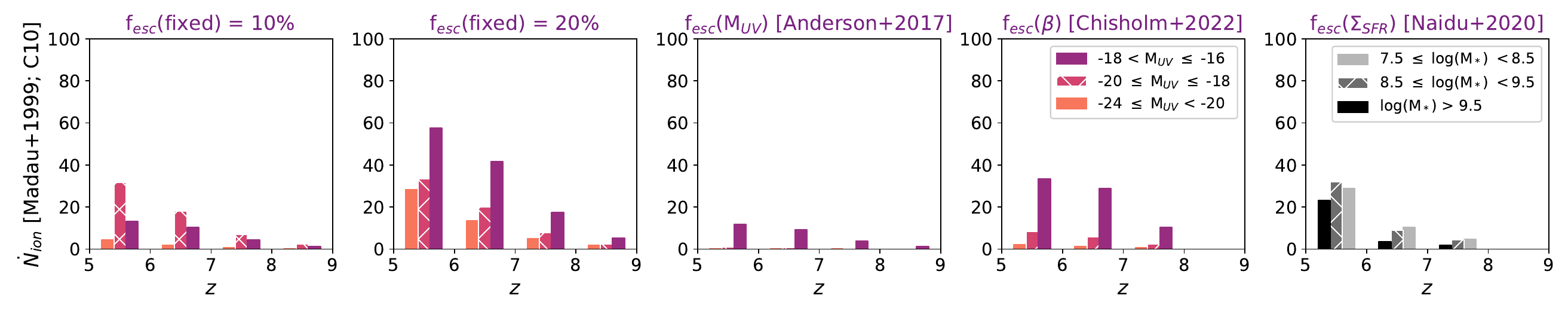}
    \caption{Percentage of cosmic rate of ionising photons being emitted per second and per unit volume, as a function of M$_{\rm{UV}}$ (four leftmost columns, coloured) and stellar mass (rightmost column, black and white). Three bins are shown per redshift in each panel, as indicated in the legends. The three  M$_{\rm{UV}}$ bins correspond to those shown in Figure~\ref{fig:MUV_mass_zbins}, where brighter galaxies tend to have higher stellar masses.
    Each row is normalised to the \citet{Madau1999} models shown in Figure~\ref{fig:cosmic_Ndot}, where the \Ndot\/ needed to maintain hydrogen ionisation at every redshift bin is set to 100\%: from top to bottom, C=1, C =3 and C=10. The columns show the different \fesc\/ prescriptions adopted earlier: constant (10 and 20\%), and varying as a function of: M$_{\rm{UV}}$ following \citet{Anderson2017}, $\beta$ as proposed in \citet{Chisholm2022}, and $\Sigma_{\rm{SFR}}$ from \citet{Naidu2020}. It can be seen that as the clumping factors increase, it becomes more difficult for galaxies to produce enough ionising photons by the end of the EoR ($z \sim 5-6$). Importantly, for 
    the calculations made convolving $\rho_{\rm{UV}}$ (four leftmost columns), the faintest  galaxies (fainter than M$_{\rm{UV}} = -20$) dominate the ionising photon budget for every \fesc\/ prescription and clumping factor. Analogously, the models on the rightmost column indicate that the intermediate and low mass bins dominate the budget, with a significant contribution from the galaxies in the lowest mass bin (log(M$_{\star}$/[M$_{\odot}$])<8.5). Therefore, despite the uncertainties in the clumping factor, we confirm that faint galaxies with low stellar masses are key agents of reionisation.}
    \label{fig:cosmic_Ndot_percentages}
\end{figure*}

\section{Implications for the reionisation of the Universe}
Recent works based on JWST observations have arrived to the conclusion that there is an overestimation of ionising photons at the EoR, leading to a so called "crisis" in the ionising photon budget \citep[see e.g., ][]{Trebitsch2022,Chakraborty2024,Munnoz2024}. In the following sections, we discuss the implications of our \xionnofesc\/ estimations for a stellar mass complete sample on the cosmic ionising budget.

\label{sec:implications_reionisation}
\subsection{Constraints on the cosmic ionising budget}
As in \cite{Simmonds2023_JADES}, we now study the implications of our findings on the ionising cosmic budget through \Ndot\/, which represents the number of ionising photons produced per volume unit. We first analyse the contributions of different M$_{\rm{UV}}$ and redshift bins to \Ndot\/, by adopting the UV luminosity functions from \cite{Bouwens2021} and some simple prescriptions for \fesc\/: constant of 10 and 20\% \citep{Ouchi2009,Robertson2013,Robertson2015}, and varying according to M$_{\rm{UV}}$ \citep{Anderson2017}. The latter was constructed using the uniform volume simulation \texttt{Vulcan}, and states the dependence: log(\fesc\/) = $(0.51\pm 0.4)$M$_{\rm{UV}} +7.3 \pm 0.08$. For \xion\/ we use the \xionnofesc\/ relations presented in Figure~\ref{fig:xion_MUV_zbins} for the star-forming sample (i.e. log(SFR$_{10}$/SFR$_{100}$) $\geq$ -1). This sample was chosen because it is representative of most galaxies in this study, accounting for >97\% of the total sample. We remind the reader that our \texttt{Prospector}-inferred ionising photon production efficiencies assume \fesc\/=0, such that \xionnofesc\/ = \xion\/$\times$(1-\fesc\/). For simplicity, in the following calculations we adopt \xion\/=\xionnofesc\/, therefore, the resulting \Ndot\/ values should be taken as lower limits. For reference, the mean \fesc\/ for the star-forming sample is $\sim3$\% (see Figure~\ref{fig:xion_fesc_remnants}), which would result in a difference in \Ndot\/ of <0.1 dex.

The resulting \Ndot\/ curves as a function of M$_{\rm{UV}}$ are shown in Figure~\ref{fig:LF_MUV}, each panel displaying a different \fesc\/ as indicated in the top left corner. As expected, the curves with fixed \fesc\/ are flatter than those derived in \cite{Simmonds2023_JADES}, a consequence of now having a stellar mass complete sample that leads to a milder \xion\/ evolution with M$_{\rm{UV}}$. Once all types of galaxies are potentially included, the importance of faint galaxies in the ionising cosmic budget is reduced. We note that our exclusion of mini-quenched galaxies does not change this conclusion, since they only account for $\lesssim 4$\% of the sample. In the case of the variable \fesc\/, the curves are still steep, due to the nature of their \fesc\/ prescriptions. On the one hand, \cite{Anderson2017} states that ionising photons can escape much more easily from faint galaxies. On the other hand, \cite{Chisholm2022} propose that galaxies with bluer UV continuum slopes ($\beta$) are likely more efficient LyC leakers. The $\beta$ in Equation~\ref{eq:fesc_Chisholm} can be expressed as a function of M$_{\rm{UV}}$ as:
\begin{equation}
    \beta = b+a\times(\text{M}_{\rm{UV}}+19.5)
\end{equation}
As described in \cite{Chisholm2022}, the coefficients $a$ and $b$ connect M$_{\rm{UV}}$ to the UV continuum slope, $\beta$, through the $\beta$-M$_{\rm{UV}}$ relations provided in \cite{Bouwens2014} for selected redshifts. On the top left panel, we also include the results if we adopt a constant log(\xion\//[Hz erg$^{-1}$]) = 25.2 \citep[based on stellar populations, as in ][]{Robertson2013}. We find that the shapes of the \Ndot\/ curves with fixed \xion\/ are  similar to the ones found in this work, albeit slightly offset towards lower \Ndot\/.

Using the \Ndot\/ curves estimated above, in Figure~\ref{fig:cosmic_Ndot} we investigate how the ionising cosmic budget evolves with redshift, given by:
\begin{equation}
    \dot{N}_{\rm{ion}} = \int\limits_{\text{M}_{\rm{UV,min}}}^{\text{M}_{\rm{UV,max}}} \rho_{\rm{UV}}(\text{M}_{\rm{UV}},z) \times \text{f}_{\rm{esc}}(\text{M}_{\rm{UV}},z) \times \xi_{\rm{ion}}(\text{M}_{\rm{UV}},z)\text{dM}_{\rm{UV}},
\end{equation}
where \Ndot\ is in units of s$^{-1}$ Mpc$^{-3}$, \xion\ is in units of Hz erg$^{-1}$, and $\rho_{\rm{UV}}$ in units of erg s$^{-1}$ Hz$^{-1}$ Mpc$^{-3}$, and the escape fraction is dimensionless. We assume that the scatter is negligible and that there is an interdependence of variables (e.g. \xion\/-M$_{\rm{UV}}$ relation). In this figure we integrate the curves shown in Figure~\ref{fig:LF_MUV} down to M$_{\rm{UV}}$ $= -16$ (open triangles), estimated by adopting the luminosity functions from \cite{Bouwens2021}. We remind the reader that our sample is UV complete down to M$_{\rm{UV}} \sim -16$, and thus, our derived relations are valid in the integrated range. We also include the luminosity densities from \cite{Sun2016}, who fit a power law to the low-mass end. To estimate \xion\/ we use the best-fit relation to our data, as shown in  Figure~\ref{fig:xion_MUV_zbins}. For consistency, we use only the star-forming galaxies (accounting for $\sim 94$\% of the total sample).

When adopting the \cite{Sun2016} luminosity functions, we only assume fixed escape fractions of 10 and 20\% (filled circles), motivated by the canonical average \fesc\/ needed for galaxies to ionise the Universe \citep{Ouchi2009,Robertson2013,Robertson2015}. As comparison, we add curves indicating the \Ndot\/ needed to maintain ionisation of hydrogen, according to the models of \cite{Madau1999}, for clumping factors of 1, 3 and 10 \citep[although see ][for a discussion of the validity of this approach at the EoR]{So2014}. A clumping factor of unity represents a uniform IGM, while larger clumping factors imply a higher number of recombinations taking place in the IGM, and thus, more ionising photons need to be emitted in order to sustain ionisation. Finally, the values found in \cite{Simmonds2023_JADES} for ELGs, are shown as a shaded hatched area. These were provided as upper limits, since they were estimated based on one important assumption: that the sample of ELGs was representative of the entire galaxy population. Indeed, in a recent work by \cite{Munnoz2024}, an excess of ionising photons inferred from JWST observations is identified in the cosmic ionising budget. With our updated stellar mass complete sample, we are now able to provide more realistic results. We find that the estimations made with our star-forming sample are consistent with those from literature \citep[e.g. ][]{Bouwens2015,Mason2015,Mason2019,Naidu2020,Rinaldi2024HAEs}.

In addition to the measurements described above, we include those found through the stellar mass function (instead of $\rho_{\rm{UV}}$). In particular, we use results from \cite{Weibel2024}, which are particularly relevant to this work since they were constructed from NIRCam observations of galaxies at $z\sim4-9$. In this scenario, \Ndot\/ can be rewritten as:
\begin{equation}
    \dot{N}_{\rm{ion}} = \int\limits_{\text{M}_{\rm{min}}}^{\text{M}_{\rm{max}}} \Phi(\text{M},z) \times \text{f}_{\rm{esc}}(\text{M},\text{SFR(M)},\text{R(M)}) \times \dot{n}_{\rm{ion}}(\text{M},z)\text{dM},
\end{equation}
where $\Phi$ is in units of Mpc$^{-3}$, \fesc\/ is dimensionless and depends on stellar mass, SFR, and size of the galaxy (through the SFR surface density, $\Sigma_{\rm{SFR}}$), and \ndot\/ is in units of s$^{-1}$. In this case, we adopt the \fesc\/ prescription from \cite{Naidu2020} due to its dependence on $\Sigma_{\rm{SFR}}$, such that: 
\begin{equation}
    \text{f}_{\rm{esc}} = min \bigg(1, 1.6_{-0.3}^{+0.3} \times \Big(\frac{\Sigma_{\rm{SFR}}}{1000 \text{M}_{\odot} \text{yr}^{-1} \text{kpc}^{-2}}\Big)^{{0.4}_{-0.1}^{+0.1}} \bigg)
\end{equation}
In order to obtain $\Sigma_{\rm{SFR}}$, we first follow the size-mass relation and coefficients (for the full sample) presented in Table 3 of \cite{Morishita2024}. We then combine the derived sizes with SFR$_{10}$ (averaged over the past 10 Myr), such that $\Sigma_{\rm{SFR}} = \frac{\text{SFR}/2}{\pi \text{R}^2}$ \citep{Shibuya2019}. After estimating $\Sigma_{\rm{SFR}}$, we compute \fesc\/$\times$\ndot\/ for our stellar mass-complete sample, for each redshift bin. By combining the best-fit relations with $\Phi$, and integrating them down to log(M$_{\star}$/[M${_\odot}$])=7.5, we finally find \Ndot\/ as a function of redshift. The values are shown as stars in Figure~\ref{fig:cosmic_Ndot}, and are in general agreement with the other variable \fesc\/ prescriptions \citep[i.e. ][]{Anderson2017,Chisholm2022}.


Although the values of \Ndot\/ are highly uncertain in the early Universe, strong constraints on \Ndot\/ have been placed by observations of the \lya\/ forest at lower redshifts \citep[e.g. ][ included in Figure~\ref{fig:cosmic_Ndot}]{Becker_Bolton2013,Gaikwad2023,Davies2024}. The \lya\/ forest refers to a series of absorption lines at wavelengths redder than \lya\/ ($\lambda_{\rm{rest-frame}} \sim 1216$ \AA\/) observed in the spectra of high redshift quasars, produced by the intervening neutral IGM. Interestingly, \Ndot\/ has been observed to flatten once reionisation has been completed (with log(\Ndot\//[s$^{-1}$ Mpc$^{-3}$])$\sim 50.8$ at $z \lesssim 5-6$). We find that, out of all the prescriptions adopted in this work, the \fesc\/ relations from \cite{Chisholm2022} can match the shape of the \lya\/ forest constraints. We remind the reader that in order to construct a stellar mass-complete sample, we have ignored all galaxies below the completeness limit (log(M$_{\star}$/[M$_{\odot}$])$\sim 7.5$). Therefore, our points represent lower limits to the cosmic ionising budget.

\subsection{Which galaxies reionised the Universe?}
In order to determine which galaxies dominate the budget of reionisation, we use the same \fesc\ prescriptions as before, but integrate their contributions in three M$_{\rm{UV}}$ bins: $-24 \leq \text{M}_{\rm{UV}} < -20$, $-20 \leq \text{M}_{\rm{UV}} \leq -18$, and $-18 < \text{M}_{\rm{UV}} \leq -16$. Figure~\ref{fig:MUV_mass_zbins} shows 
that M$_{\rm{UV}}$ and stellar mass are correlated (albeit with a large scatter),
where brighter galaxies are more massive (and vice-versa), such that each UV luminosity bin also loosely describes a mass bin in our sample. In the four leftmost columns of  Figure~\ref{fig:cosmic_Ndot_percentages}  we adopt these M$_{\rm{UV}}$ luminosity bins and show the relative contribution these bins have in \Ndot\/ for each redshift bin above 5 \citep[where observational studies agree the EoR has ended, e.g. ][]{Keating2020,Yang2020,Zhu2024}. In the rightmost column we show the contributions of different stellar mass bins to the ionising budget, by adopting the stellar mass functions of \cite{Weibel2024} and the \fesc\/ prescription from \cite{Naidu2020} (stars in Figure~\ref{fig:cosmic_Ndot}).
Each row has been normalised to a different clumping factor using the models from \citep{Madau1999}. We find that if the IGM is uniform (C=1), then galaxies produce enough ionising radiation in order to sustain hydrogen ionisation, independent of the \fesc\/ prescription. As the clumping factor increases, it becomes more difficult for galaxies to ionise the Universe by redshift 5. However, most importantly, we discover that for every clumping factor, \fesc\/ assumption, and redshift bin, the fainter galaxies (with M$_{\rm{UV}} >= -20$), and galaxies with low and intermediate stellar masses (log(M$_{\star}$/[M$_{\odot}$])<9.5) dominate the cosmic ionising budget. In agreement with the results presented in \cite{Seeyave2023}, based the First Light and Reionization Epoch Simulations \citep[FLARES; ][]{Lovell2021, Vijayan2021}, and the conclusions reached by forward modelling JWST analogues from the SPHINX simulation, presented in \cite{Choustikov2024}.

Therefore, in this work, we confirm that faint low-mass galaxies with bursty star formation have in general enhanced \xion\/ compared to massive galaxies and/or galaxies without recent star formation, in agreement with \cite{Simmonds2023_JADES}. However, when taking into account the full galaxy population, their contribution is less extreme as might have been thought previously, resolving any potential crisis in the ionising photon budget of the Universe \citep[e.g., ][]{Pahl2024,Trebitsch2022,Chakraborty2024}. Finally, by adopting the relations from \cite{Chisholm2022}, we are able to reconcile our results with the constraints provided by observations of the \lya\/ forest, especially in the flattening of \Ndot\/ that has been observed at lower redshifts ($z \lesssim 6$). In summary, we have shown that galaxies produce enough ionising photons to ionise the Universe by $z \approx 5-6$, without creating a nonphysical excess of ionising photons in the cosmic budget.

\section{Caveats and limitations}
\label{sec:caveats}
There are a few important limitations to our method, which we now describe. Firstly, since our method relies on SED fitting, and includes the assumption of certain stellar populations, we are incapable of detecting and appropriately fitting extreme populations. In particular, we assumed a Chabrier IMF, with a maximum stellar mass of 100 M$_{\odot}$. This choice was motivated by the size of the sample, in an effort to chose a representative IMF and stellar populations. However, if there are extreme objects in our sample, we might be missing them. For example, \cite{Cameron2023} have found tentative evidence for a top heavy IMF at $z \sim 6$ although, \cite{Tacchella2024} find an alternative explanation that does not require to invoke exotic stellar populations. If the IMF evolves with redshift, then our choice of a Chabrier IMF would affect our results (including the derived stellar masses). We circumvent this issue to some extent by selecting only the galaxies that were fit with a reduced $\chi^2 < 1$, but we note that by doing so, might have lost information on sources that cannot be reproduced by our models.

In the same vein, our modelling does not include a prescription for AGN. JWST has recently unveiled a hidden population of AGN at high redshifts \citep{Juodzbalis2023,Madau2024,Maiolino2024,Ubler2024}. Underestimating the AGN contribution can lead to a systematic overestimation of stellar mass and SFRs by SED fitting codes \citep{Buchner2024}. Unfortunately, accurately identifying which galaxies in our sample host AGN is not trivial, as demonstrated by \cite{Wasleske2024}, who compare different techniques used to select AGN in dwarf galaxies (M$_* \leq 10^{9.5}$M$_{\odot}$). They find that  any single diagnostic can retrieve at most half of the AGN sample, and most importantly for this work, the AGN identification is least effective when considering photometry alon. Therefore, quantifying the contribution of AGN to our sample is far from the scope of this work. As such, we report our results and caution that our sample might contain some AGN that can be mimicked by stellar emission. 

Another important point is that our work relies heavily on photometric redshift measurements. Specifically, our \texttt{Prospector}-inferred redshifts use \texttt{EAzY} redshifts as priors. The latter has been proven to provide accurate results for large samples using JADES NIRCam photometry in GOODS-S \citep{Rieke2023}. Moreover, in this work we compare them (when possible) to spectroscopic redshifts compiled from literature. We find a good --but not perfect- agreement, suggesting that our results are in general accurate but likely contain a few outliers where the $z_{\rm{phot}}$ is unreliable. We stress that these outliers would be of the order of at most a few percent. As with the stellar populations, our selection of galaxies (fitted with a reduced $\chi^2 \leq 1$) ensures that the photometry is well represented by the best fit models. Moreover, the addition of HST observations aids in constraining the photometric redshifts. Finally, we do not expect cosmic variance to play an important factor in our results, but plan in the future to perform a similar study in GOODS-N.

In summary, whereas there are intrinsic limitations to our methods, our results are --as much as possible-- accurate and representative of the galaxy population in GOODS-S (at $3 \leq z \leq 9$).  

\section{Conclusions}
\label{sec:conclusions}
We use JWST NIRCam photometry to build a sample of 14652 galaxies at $3 \leq z \leq 9$, 1640 of them with spectroscopic redshifts from literature. We infer their properties using the SED fitting code \texttt{Prospector}, finding two distinct populations of galaxies which can be separated by their burstiness (delimited by log(SFR$_{10}$/SFR$_{100}$) = -1). We  call these populations star forming and mini-quenched, and note that the mini-quenched galaxies only account for $<3\%$ of the total sample. Within the mini-quenched population, we find an interesting subsample with tentative evidence of LyC leakage (through the UV continuum slope $\beta$). These galaxies populate a similar parameter space as the remnant leakers from \cite{Katz2023}. Future spectroscopic follow-ups will be necessary to confirm or refute this hypothesis. Our main findings can be summarised as follows.\\

We find that \xionnofesc\/ increases for fainter galaxies with burstier SFHs, in agreement with previous studies, albeit with a milder evolution with redshift. The latter is explained by the nature of our sample, and that previous studies were biased towards galaxies with strong emission lines and/or LAEs. The evolution of \xionnofesc\/ with $z$ for the more representative star-forming sample is:
\begin{equation*}
    \log (\text{\xionnofesc\/(z)}) = (-0.001\pm 0.004)z + (25.294\pm 0.017)
\end{equation*}

\noindent The 2-dimensional fit that accounts for the change of \xion\ with M$_{\rm{UV}}$ and redshift, for the same sample is given by:
\begin{multline*}
\log(\xi_{\rm{ion,0}} (z,\text{M}_{\rm{UV}})) = \\ 
    (0.003\pm 0.003)z + (-0.018\pm 0.003)\text{M}_{\rm{UV}} + (25.984\pm 0.053)
\end{multline*}

To study the contribution of the galaxies in this study to reionisation, we convolve the star-forming relations (which represent $>97\%$ of the total sample), with luminosity functions from literature. We find that galaxies, which are detected with JWST, can ionise the Universe by the end of the EoR, if we assume the AGN contribution is minor. In particular, assuming a fixed escape fraction, we find that galaxies fainter than M$_{\rm{UV}} = -20$ contribute similar amounts of ionising photons (see Figure~\ref{fig:LF_MUV}), and that galaxies in the range of M$_{\rm{UV}} = -20$ to $-16$ dominate the budget of reionisation at every redshift bin studied in this work (see Figure~\ref{fig:cosmic_Ndot_percentages}). With our stellar mass complete sample, our predictions do not overestimate \Ndot\ for galaxies at $z > 8$ \citep[see][]{Munnoz2024}. We note that if we extrapolate our trends to fainter magnitudes (to M$_{\rm{UV}}$ of -14 or -12), the Universe can be reionised with lower escape fractions. Promisingly, by adopting the relation of \fesc\/ with M$_{\rm{UV}}$ presented in \cite{Chisholm2022}, we can conciliate our results regarding the ionising cosmic budget with the constraints obtained through observations of the \lya\/ forest. 

\section*{Acknowledgements}
The JADES Collaboration thanks the Instrument Development Teams and the instrument teams at the European Space Agency and the Space Telescope Science Institute for the support that made this program possible. We also thank our program coordinators at 
STScI for their help in planning complicated parallel observations. We would like to acknowledge the FRESCO team, lead by Pascal Oesch, for developing their observing program with a zero-exclusive-access period.

CS thanks Sarah Bosman, Frederick Davies, James Leftley, and Arjen van der Wel for the insightful discussions. CS, W.B., RM, \& JW acknowledge support by the Science and Technology Facilities Council (STFC) and by the ERC through Advanced Grant number 695671 ‘QUENCH’, and by the UKRI Frontier Research grant RISEandFALL. RM also acknowledges funding from a research professorship from the Royal Society. ST acknowledges support by the Royal Society Research Grant G125142. BER, BDJ, PAC \& YZ acknowledge support from the NIRCam Science Team contract to the University of Arizona, NAS5-02015. BER also acknowledges support by the JWST Program 3215. DP acknowledges support by the Huo Family Foundation through a P.C. Ho PhD Studentship. This research is supported in part by the Australian Research Council Centre of Excellence for All Sky Astrophysics in 3 Dimensions (ASTRO 3D), through project number CE170100013. AJB, JC \& GCJ acknowledge funding from the "FirstGalaxies" Advanced Grant from the European Research Council (ERC) under the European Union’s Horizon 2020 research and innovation programme (Grant agreement No. 789056). SC acknowledges support by European Union’s HE ERC Starting Grant No. 101040227 - WINGS. ECL acknowledges support of an STFC Webb Fellowship (ST/W001438/1). IL is supported by the National Science Foundation Graduate Research Fellowship under Grant No. 2137424. H{\"U} gratefully acknowledges support by the Isaac Newton Trust and by the Kavli Foundation through a Newton-Kavli Junior Fellowship. NCV acknowledges support from the Charles and Julia Henry Fund through the Henry Fellowship. The research of CCW is supported by NOIRLab, which is managed by the Association of Universities for Research in Astronomy (AURA) under a cooperative agreement with the National Science Foundation.

\section*{Data Availability}
The data underlying this article will be shared on reasonable request to the corresponding author.



\bibliographystyle{mnras}
\bibliography{bib} 



\appendix

\section{Goodness of fits}
\label{app:chi}
\begin{figure*}
    \includegraphics[width=2\columnwidth]{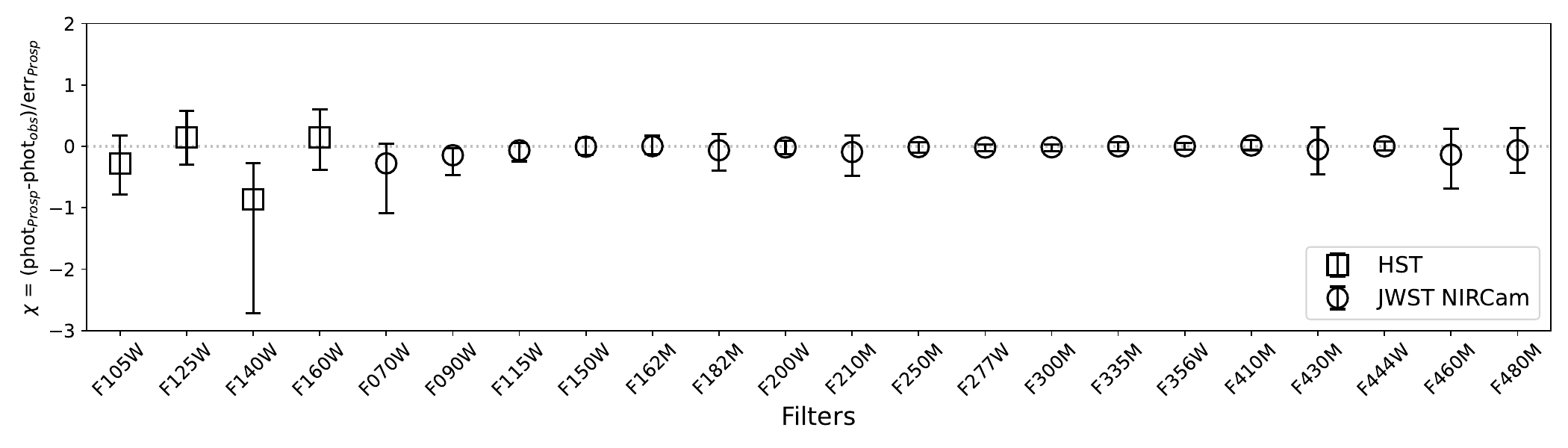}
    \caption{Comparison between the modelled and observed HST (squares) and JWST NIRCam (circles) photometry for our stellar mass complete sample. $\chi$ scatters around zero with mostly symmetric error bars.}
    \label{fig:chi_filters}
\end{figure*}

\begin{figure*}
    \includegraphics[width=2\columnwidth]{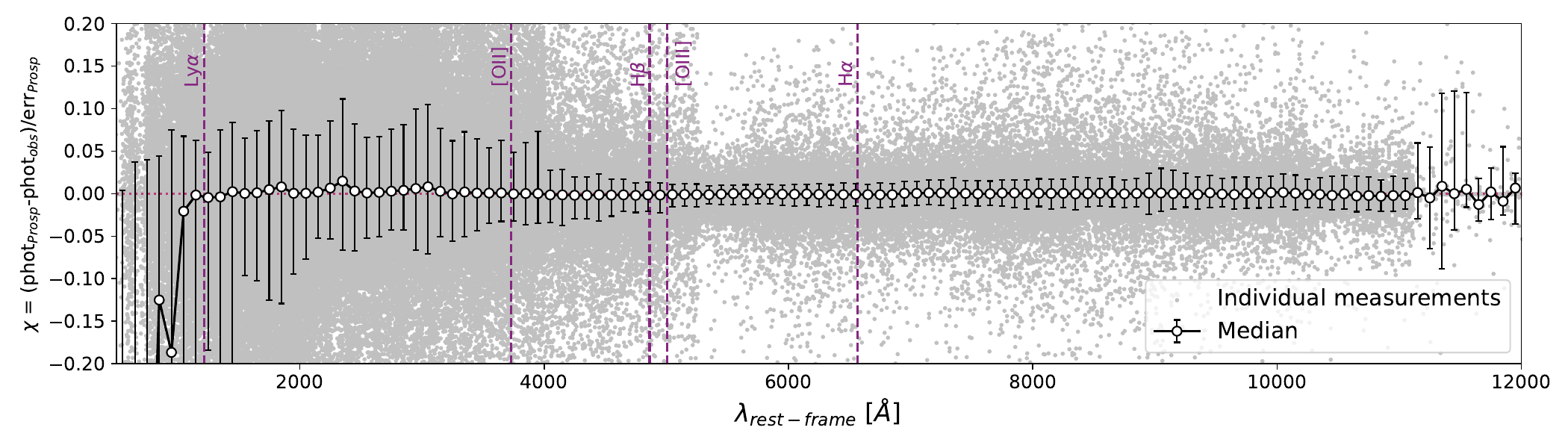}
    \caption{Same as Figure~\ref{fig:chi_filters} but at rest-frame wavelengths. The grey dots show the individual measurements, while the white circles and error bars show the median and errors when adopting bins of width 100\AA\/. For comparison, the \lya\/, \oii\/, \hb\/, \oiii\/, and \ha\/ emission wavelengths are shown as vertical dashed lines. It can be seen that the median values scatter around $\chi = 0$, with symmetrical error bars. Moreover, we find no evidence of emission line fluxes being over (or under) estimated by \texttt{Prospector}.}
    \label{fig:chi_restframe}
\end{figure*}

Here we present a comparison between the modelled and observed photometry,  defined as the difference between them, divided by the error in the modelled points. The median and errors are shown in Figure~\ref{fig:chi_filters}, where we exclude the HST bands F435W, F606W, F775W, F814W and F850LP. The observations in these filters are highly uncertain for our sample, yielding $\chi$ values well outside of the bounds of the figures. $\chi$ scatters around zero, with error bars that are mostly symmetric. As expected, the deep NIRCam photometric set is, in general, better  represented by the \texttt{Prospector} best-fit models.  Figure~\ref{fig:chi_restframe} shows the same comparison but at rest-frame wavelengths (where the \texttt{Prospector} photometric redshift has been adopted). The measurements have been binned into wavelength bins of width 100\AA\/, and the median values and errors are shown as white circles and error bars. The latter scatter around zero, and most importantly, do not show indication of emission line fluxes being either over or underestimated by \texttt{Prospector}.  The photometric offsets for each band, defined as the ratio between the observed and modelled fluxes, are given in Table~\ref{tab:offsets}.   

\begin{table}
    \caption{Photometric offsets defined as the ratio between the observed and modelled photometry. \textsl{Column 1:} name of filter. \textsl{Colum 2:} median offset and errors, given by the 16$^{\rm{th}}$ and 84$^{\rm{th}}$ percentiles.}
    \centering
    \begin{tabular}{c|c}
    \hline
    \noalign{\smallskip}
    Band & Offset \\ 
    \noalign{\smallskip}
    \hline
    \noalign{\smallskip}
    F435W & 2.8 (+86.92, -2.54)\\
    F606W & 2.0 (+50.04, -1.81)\\
    F775W & 1.50 (+16.86, -1.32)\\
    F814W & 1.34 (+12.06, -1.17)\\
    F850LP & 1.44 (+9.22, -1.25)\\
    F105W & 1.17 (+7.21, -1.01)\\
    F125W & 1.08 (+6.71, -0.93)\\
    F140W & 1.96 (+10.73, -1.70)\\
    F160W & 1.20 (+7.16, -1.04)\\
    F070W & 1.69 (+17.41, -1.47)\\
    F090W & 1.51 (+9.14, -1.27)\\
    F115W & 1.44 (+7.47, -1.19)\\
    F150W & 1.40 (+7.09, -1.16)\\
    F162M & 1.63 (+8.89, -1.36)\\
    F182M & 1.48 (+8.50, -1.25)\\
    F200W & 1.36 (+7.77, -1.14)\\
    F210M & 1.45 (+9.20, -1.24)\\
    F250M & 1.68 (+9.11, -1.42)\\
    F277W & 1.39 (+7.18, -1.16)\\
    F300M & 1.73 (+10.06, -1.44)\\
    F335M & 1.60 (+8.50, -1.35)\\
    F356W & 1.43 (+7.37, -1.19)\\
    F410M & 1.53 (+8.08, -1.30)\\
    F430M & 1.50 (+8.14, -1.30)\\
    F444W & 1.50 (+7.35, -1.25)\\
    F460M & 1.42 (+8.04, -1.22)\\
    F480M & 1.22 (+6.89, -1.04)\\
    \hline
    \end{tabular}
    \label{tab:offsets}
\end{table}

\section{Shape of redshift and stellar mass distributions}
\label{app:corner}
\begin{figure}
	\includegraphics[width=\columnwidth]{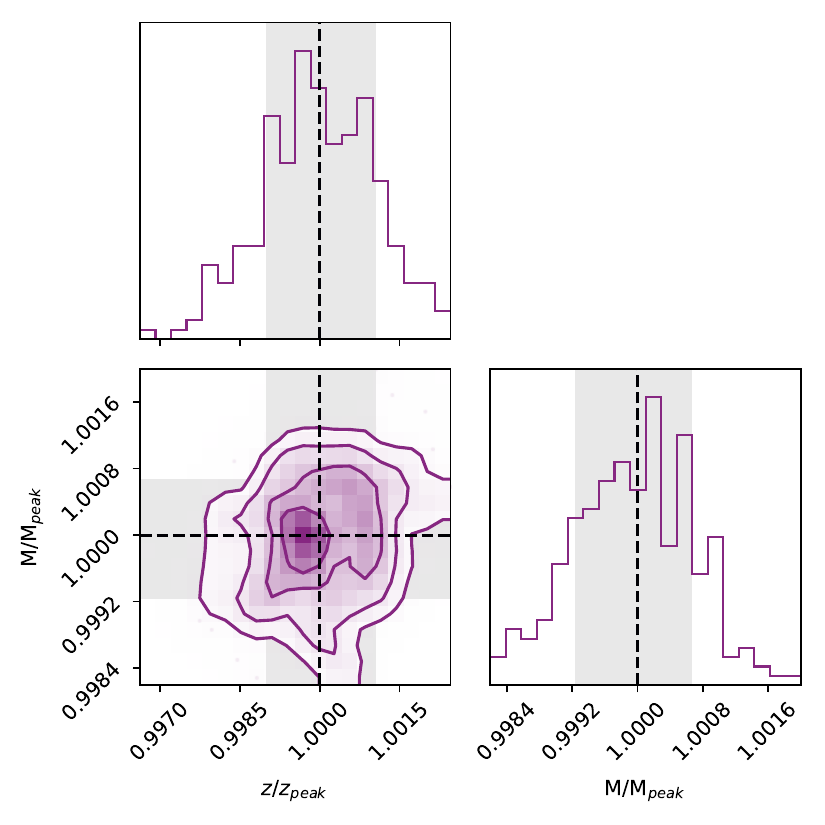}
    \caption{Corner plot showing the main properties of interest in the stellar mass completeness estimation: redshift and stellar mass. The values have been normalised to the 50$^{\rm{th}}$ percentile (dashed vertical lines) for each galaxy, in order to understand the general shape of the posteriors in our sample. The grey areas show the 16$^{\rm{th}}$ and 84$^{\rm{th}}$ ranges. We find no significant signatures of  asymmetry.}
    \label{fig:corner}
\end{figure}

Figure~\ref{fig:corner} shows the mean shape of the main properties used to estimate the stellar mass completeness in Section~\ref{section:completeness}. The redshift and stellar masses have been divided by the 50$^{\rm{th}}$ percentile value for each galaxy. We find no significant signs of asymmetry in the posteriors for our sample, and the values are in general well constrained by the fitting routine.

\section{Information gained after SED-fitting}
\label{app:IG}

\begin{figure*}
	\includegraphics[width=\columnwidth]{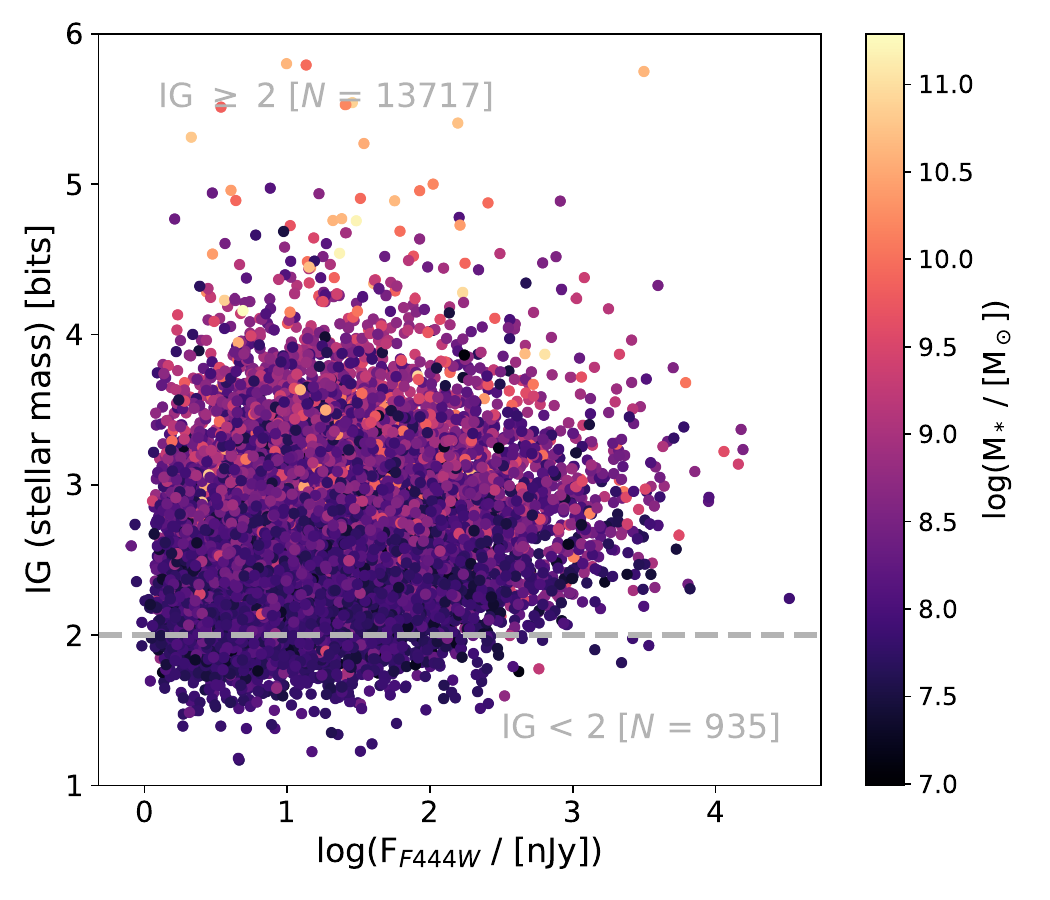}
 \includegraphics[width=\columnwidth]{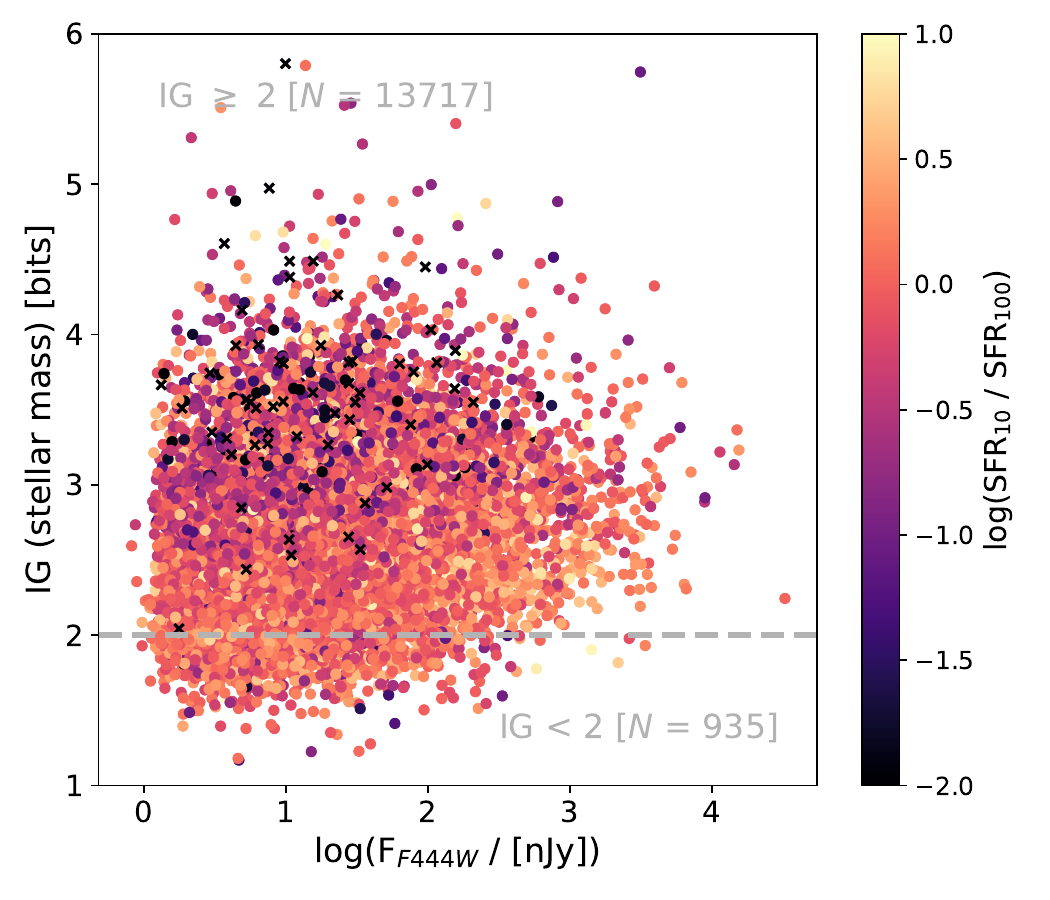}
    \caption{Kullback-Leibler information gain as a function of flux in the F444W band. A higher flux leads in general to a higher IG. However, all of the galaxies in our sample have IG > 1, with $\sim 94$\% above IG = 2. \textsl{Left panel:} colour-coded by stellar mass. \textsl{Right panel:} colour-coded by burstiness of their SFH, defined as SFR$_{10}$/SFR$_{100}$. The crosses show galaxies with no recent star formation (SFR$_{10}$ = 0).}
    \label{fig:IG}
\end{figure*}

\begin{figure*}
	\includegraphics[width=\columnwidth]{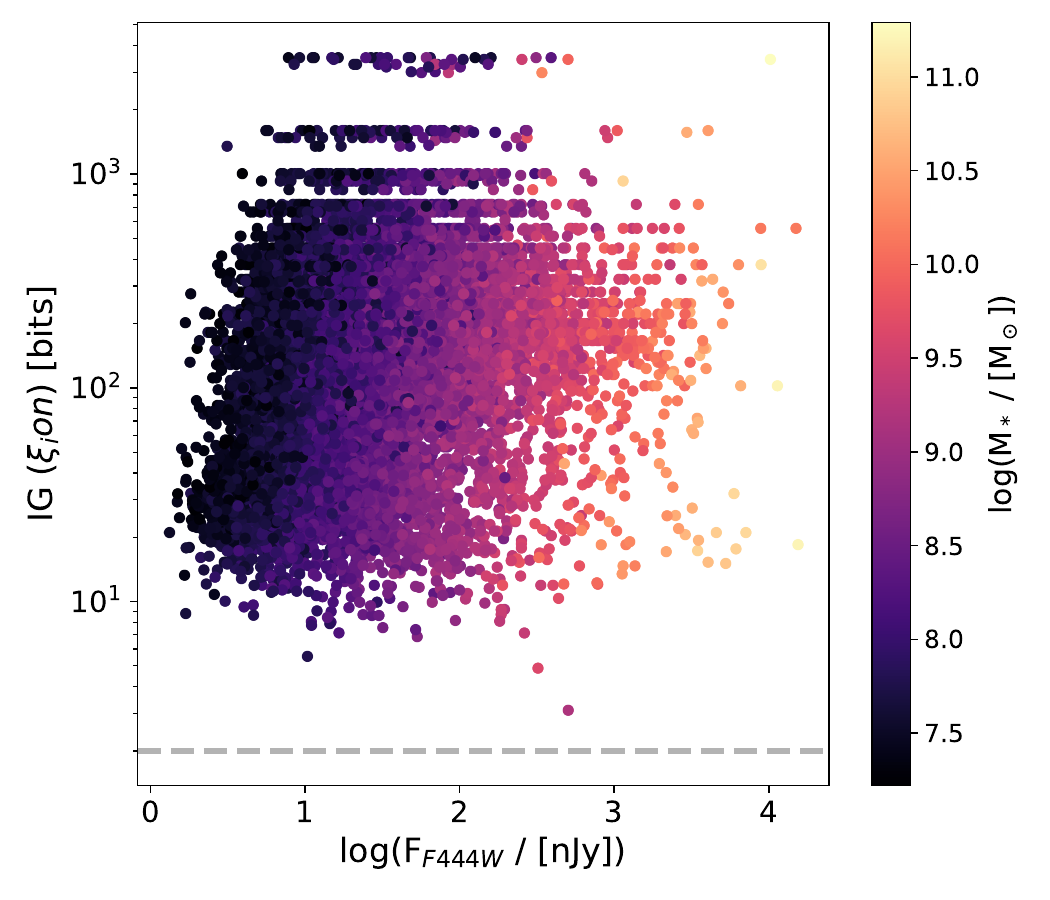}
 \includegraphics[width=\columnwidth]{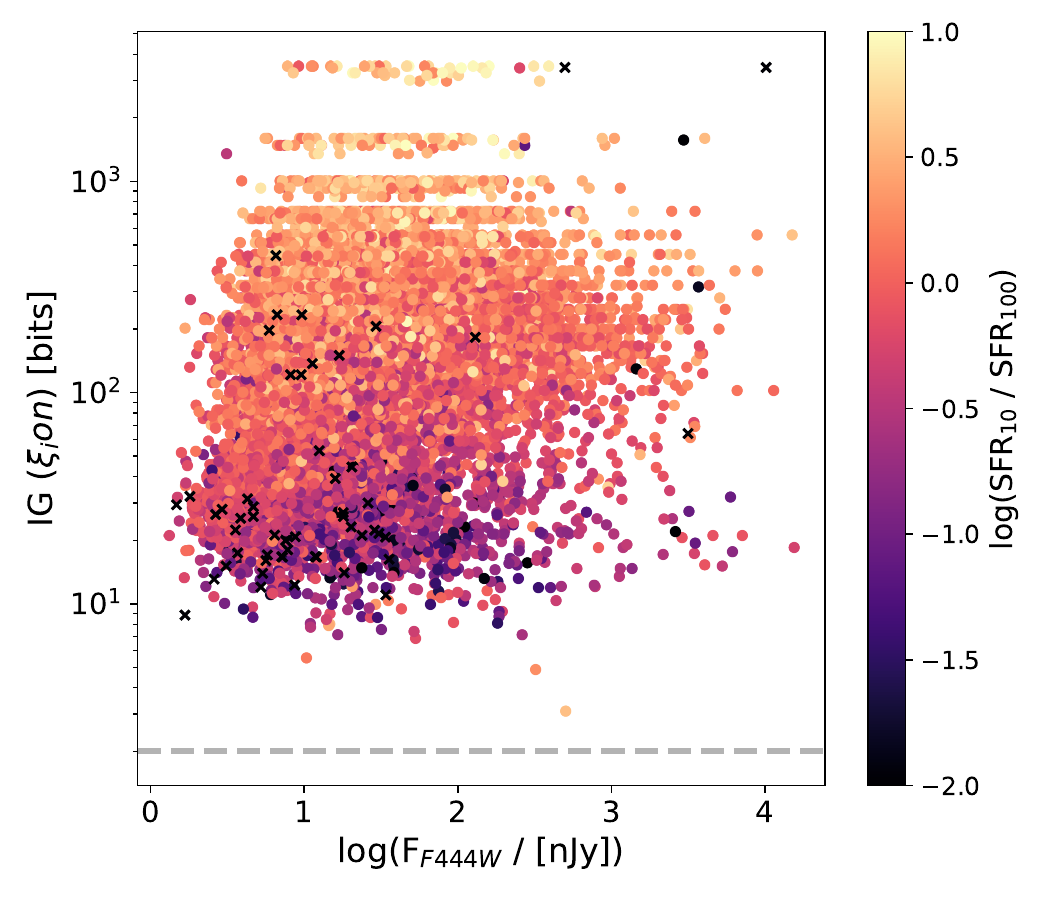}
    \caption{Same as Figure~\ref{fig:IG} but for the ionising photon production efficiency, \xionnofesc\/. The information gain tends to increase towards galaxies with burstier SFHs, however, we note that all galaxies in our stellar mass complete sample have a high information gain (well above IG = 2, shown as a dashed horizontal line).}
    \label{fig:IG_xion}
\end{figure*}

In order to measure how well \texttt{Prospector} can constrain stellar masses and ionising photon production efficiencies in our sample, we use the Kullback-Leibler definition of information gain (IG), given by:
\begin{equation}
    \text{IG} = \int \text{Posterior}(p) \times \log_2 \frac{\text{Posterior}(p)}{\text{Prior}(p)} \text{d}p \text{[bits]},
\end{equation}
where $p$ represents the parameter of choice: either $\log\text{M}_{\star}$ or $\log\xi_{\rm{ion}}$.
IG is a way of measuring the difference between the prior and posterior, in units of bits: an IG = 0 means the prior is equal to the posterior and no information was gained, whereas a higher value of IG indicates a larger amount of information was gained in the fitting routine. Following the criteria from \cite{Simmonds2018}: 
\begin{itemize}
    \item IG < 1: little-to-no information was gained
    \item 1 $\leq$ IG $\leq$ 2: some information was gained
    \item IG > 2: parameter is constrained
\end{itemize}
Figures~\ref{fig:IG} and ~\ref{fig:IG_xion} show the IG for stellar mass and \xionnofesc\/, respectively, for our stellar mass complete sample. They are shown as a function of the flux in the F444W filter, colour-coded by stellar mass (left) and by burstiness (right). As expected, IG(stellar mass) is highest for brighter massive galaxies, and is lowest for fainter lower-mass galaxies. Importantly, all values are above 1 (the majority lie above 2) and the information gained in stellar mass does not depend on the existence of recent star formation, or strong emission lines. In the case of \xionnofesc\/, we find a trend of increasing IG with burstiness, but emphasise that all measurements are highly constrained (with IG $\gg 2$). Promisingly, most mini-quenched galaxies discussed in this work have a large IG. This is due to the richness of the photometric dataset, which allows to constrain SED shapes even in cases with no obvious emission lines. 

\section{Best-fit parameters for relations of ionising properties with observed UV magnitude}
\label{app:tables}
For readability, the relations of \xionnofesc\/ and \ndot\/, as a function of M$_{\rm{UV}}$, presented in Figures~\ref{fig:xion_MUV_zbins}
and~\ref{fig:nion_MUV_zbins}, are given in Table~\ref{tab:bestfit_xion} and~\ref{tab:bestfit_nion}, respectively.

\begin{table*}
    \caption{Best-fit parameters for log(\xionnofesc\/)-M$_{\rm{UV}}$ relations, with form log(\xionnofesc\/)$= \alpha $M$_{\rm{UV}}+$log(\xion\/$_{\rm{,int}}$) shown in Figure~\ref{fig:xion_MUV_zbins}, where \xion\/$_{\rm{,int}}$ is the intercept of the relation. \textsl{Column 1:} redshift bin. \textsl{Column 2,3:} slope and \xionnofesc\/ normalisation for the star-forming (SF) sample. \textsl{Column 4,5:} slope and \xionnofesc\/ normalisation for the mini-quenched (MQ) sample. For $z\geq7$ we only provide relations for the star-forming sample, since there are not enough mini-quenched galaxies to obtain a reliable fit.}
    \centering
    \begin{tabular}{c|c|c|c|c}
    \hline
    \noalign{\smallskip}
    Redshift & $\alpha$ & log(\xion\/$_{,0}$ / [Hz erg$^{-1}$]) & $\alpha$ & log(\xion\/$_{,0}$ / [Hz erg$^{-1}$]) \\ 
      & [SF] & [SF] & [MQ] & [MQ] \\
    \noalign{\smallskip}
    \hline
    \noalign{\smallskip}
         $3 \leq z \leq 4$ & $-0.01\pm0.00$ & $25.20\pm0.07$ & $-0.06\pm0.06$ & $23.23\pm1.04$ \\
         $4 < z \leq 5$ & $-0.02\pm0.01$ & $24.89\pm0.11$ & $-0.02\pm0.04$ & $23.95\pm0.76$ \\
         $5 < z \leq 6$ & $-0.05\pm0.01$ & $24.32\pm0.12$ & $0.03\pm0.04$ & $24.74\pm0.67$ \\
         $6 < z \leq 7$ & $-0.03\pm0.02$ & $24.88\pm0.27$ & $0.01\pm0.04$ & $24.51\pm0.77$ \\
         $7 < z \leq 8$ & $-0.05\pm0.02$ & $24.51\pm0.32$ & - & - \\
         $8 < z \leq 9$ & $-0.12\pm0.05$ & $23.28\pm0.94$ & - & - \\
    \hline
    \end{tabular}
    \label{tab:bestfit_xion}
\end{table*}

\begin{table*}
    \caption{Best-fit parameters for log(\ndot\/)-M$_{\rm{UV}}$ relations, with form log(\ndot\/)$= \alpha $M$_{\rm{UV}}+$log(\ndot\/$_{,0}$) shown in Figure~\ref{fig:nion_MUV_zbins}. \textsl{Column 1:} redshift bin. \textsl{Column 2,3:} slope and \ndot\/ normalisation for the star-forming (SF) sample. \textsl{Column 4,5:} slope and \ndot\/ normalisation for the mini-quenched (MQ) sample. For $z\geq7$ we only provide relations for the star-forming sample, since there are not enough mini-quenched galaxies to obtain a reliable fit.}
    \centering
    \begin{tabular}{c|c|c|c|c}
    \hline
    \noalign{\smallskip}
    Redshift & $\alpha$ & log(\ndot\/$_{,0}$ / [s$^{-1}$]) & $\alpha$ & log(\ndot\/$_{,0}$ / [s$^{-1}$]) \\ 
      & [SF] & [SF] & [MQ] & [MQ] \\
    \noalign{\smallskip}
    \hline
    \noalign{\smallskip}
         $3 \leq z \leq 4$ & $-0.38\pm0.01$ & $46.67\pm0.16$ & $-0.47\pm0.08$ & $43.83\pm1.38$ \\
         $4 < z \leq 5$ & $-0.40\pm0.01$ & $46.24\pm0.20$ & $-0.41\pm0.06$ & $44.78\pm1.02$ \\
         $5 < z \leq 6$ & $-0.43\pm0.01$ & $45.54\pm0.21$ & $-0.40\pm0.03$ & $44.97\pm0.61$ \\
         $6 < z \leq 7$ & $-0.40\pm0.02$ & $46.17\pm0.31$ & $-0.39\pm0.04$ & $45.13\pm0.83$ \\
         $7 < z \leq 8$ & $-0.40\pm0.02$ & $46.29\pm0.43$ & - & - \\
         $8 < z \leq 9$ & $-0.45\pm0.09$ & $45.32\pm1.71$ & - & - \\
    \hline
    \end{tabular}
    \label{tab:bestfit_nion}
\end{table*}


\bsp	
\label{lastpage}
\end{document}

%% file: Table1_excerpt.dat
JADES-GS+53.14866-27.77800 & 3.31$^{+0.19}_{-0.07}$ & 7.85$^{+0.07}_{-0.14}$ & 25.23$^{+0.25}_{-0.10}$ & 53.03$^{+0.19}_{-0.30}$ & 0.04 & -14.65$^{+0.88}_{-0.54}$\\
JADES-GS+53.12095-27.87018 & 3.33$^{+0.10}_{-0.07}$ & 8.08$^{+0.09}_{-0.14}$ & 25.37$^{+0.13}_{-0.08}$ & 53.43$^{+0.19}_{-0.21}$ & -0.13 & -17.39$^{+0.25}_{-0.11}$\\
JADES-GS+53.11867-27.78184 & 3.74$^{+0.02}_{-0.12}$ & 7.93$^{+0.13}_{-0.19}$ & 25.27$^{+0.21}_{-0.21}$ & 53.07$^{+0.27}_{-0.16}$ & 0.05 & -17.51$^{+0.08}_{-0.08}$\\
JADES-GS+53.08300-27.72764 & 3.13$^{+0.20}_{-0.24}$ & 8.23$^{+0.16}_{-0.23}$ & 25.28$^{+0.18}_{-0.24}$ & 52.93$^{+0.19}_{-0.19}$ & -0.07 & -16.60$^{+0.36}_{-0.30}$\\
JADES-GS+53.04080-27.92276 & 3.22$^{+0.21}_{-0.38}$ & 8.30$^{+0.11}_{-0.17}$ & 25.30$^{+0.17}_{-0.25}$ & 52.95$^{+0.19}_{-0.25}$ & -0.02 & -16.89$^{+0.49}_{-0.30}$\\
JADES-GS+53.09258-27.82776 & 3.10$^{+0.05}_{-0.03}$ & 9.01$^{+0.11}_{-0.17}$ & 25.35$^{+0.07}_{-0.10}$ & 53.90$^{+0.18}_{-0.17}$ & 0.30 & -18.25$^{+0.14}_{-0.11}$\\
JADES-GS+53.12590-27.74766 & 3.99$^{+1.71}_{-0.10}$ & 7.76$^{+0.28}_{-0.22}$ & 25.36$^{+0.10}_{-0.16}$ & 52.91$^{+0.46}_{-0.19}$ & 0.11 & -16.76$^{+0.34}_{-0.72}$\\
JADES-GS+53.13889-27.88894 & 3.36$^{+0.06}_{-0.15}$ & 7.61$^{+0.18}_{-0.28}$ & 25.57$^{+0.08}_{-0.04}$ & 53.43$^{+0.13}_{-0.24}$ & 0.46 & -16.04$^{+0.40}_{-0.25}$\\
JADES-GS+53.22062-27.80017 & 3.51$^{+0.19}_{-0.39}$ & 8.13$^{+0.18}_{-0.35}$ & 25.32$^{+0.19}_{-0.31}$ & 53.00$^{+0.33}_{-0.56}$ & 0.02 & -16.61$^{+0.77}_{-0.44}$\\
JADES-GS+53.10416-27.76468 & 3.22$^{+0.43}_{-0.55}$ & 7.91$^{+0.15}_{-0.25}$ & 25.16$^{+0.29}_{-0.62}$ & 52.18$^{+0.60}_{-0.93}$ & -0.15 & -14.52$^{+1.45}_{-0.99}$\\
JADES-GS+53.10666-27.85285 & 3.70$^{+0.06}_{-0.01}$ & 8.69$^{+0.25}_{-0.05}$ & 25.34$^{+0.12}_{-0.18}$ & 54.11$^{+0.11}_{-0.05}$ & 0.23 & -19.37$^{+0.05}_{-0.05}$\\
JADES-GS+53.14954-27.83692 & 3.47$^{+0.18}_{-0.20}$ & 8.44$^{+0.07}_{-0.17}$ & 24.71$^{+0.38}_{-0.62}$ & 52.42$^{+0.39}_{-0.59}$ & -0.69 & -17.37$^{+0.25}_{-0.16}$\\
JADES-GS+53.13434-27.87952 & 3.63$^{+0.06}_{-0.10}$ & 9.23$^{+0.08}_{-0.55}$ & 25.20$^{+0.17}_{-0.20}$ & 54.20$^{+0.13}_{-0.23}$ & 0.06 & -19.42$^{+0.06}_{-0.07}$\\
JADES-GS+53.07363-27.80199 & 3.79$^{+0.16}_{-0.09}$ & 8.21$^{+0.13}_{-0.16}$ & 25.06$^{+0.29}_{-0.39}$ & 52.51$^{+0.38}_{-0.43}$ & -0.30 & -16.48$^{+0.29}_{-0.24}$\\
JADES-GS+53.11274-27.86907 & 3.57$^{+0.20}_{-0.55}$ & 7.77$^{+0.15}_{-0.25}$ & 25.27$^{+0.20}_{-0.34}$ & 52.46$^{+0.34}_{-0.46}$ & -0.15 & -15.69$^{+0.59}_{-0.46}$\\
JADES-GS+53.06192-27.71806 & 3.09$^{+0.04}_{-0.02}$ & 8.01$^{+0.04}_{-0.07}$ & 25.45$^{+0.16}_{-0.02}$ & 53.76$^{+0.09}_{-0.14}$ & 0.50 & -16.01$^{+0.57}_{-0.12}$\\
JADES-GS+53.11460-27.77632 & 3.70$^{+2.91}_{-0.27}$ & 7.87$^{+0.10}_{-0.28}$ & 25.54$^{+0.12}_{-0.19}$ & 53.18$^{+0.98}_{-0.39}$ & 0.26 & -15.97$^{+0.82}_{-1.26}$\\
JADES-GS+53.12350-27.89819 & 3.06$^{+0.10}_{-0.25}$ & 8.51$^{+0.14}_{-0.20}$ & 25.24$^{+0.15}_{-0.52}$ & 53.22$^{+0.25}_{-0.46}$ & -0.09 & -18.00$^{+0.19}_{-0.13}$\\
JADES-GS+53.11707-27.76709 & 3.09$^{+0.03}_{-0.02}$ & 7.79$^{+0.15}_{-0.38}$ & 25.49$^{+0.08}_{-0.08}$ & 53.02$^{+0.34}_{-0.24}$ & 0.33 & -15.56$^{+0.50}_{-0.25}$\\
JADES-GS+53.09145-27.91365 & 3.51$^{+0.14}_{-0.18}$ & 8.26$^{+0.17}_{-0.29}$ & 25.24$^{+0.14}_{-0.27}$ & 52.95$^{+0.24}_{-0.29}$ & -0.10 & -17.28$^{+0.18}_{-0.19}$\\